\documentclass[a4paper,fleqn,usenatbib]{mnras}
\usepackage{amsmath}
\usepackage{amssymb}

\usepackage{txfonts}

\usepackage[T1]{fontenc}
\usepackage{ae,aecompl}

\usepackage{supertabular}
\usepackage{graphicx}

\newcommand{\scinote}[1]{\ensuremath{\times 10^{#1}}}
\newcommand{\expectation}[1]{\ensuremath{\left< {#1} \right>}}
\newcommand{\update}[1]{{\bfseries #1}}
\renewcommand{\update}[1]{#1} 
\newcommand{\secondupdate}[1]{{\bfseries #1}}
\renewcommand{\secondupdate}[1]{#1} 

\def \figfolder {paper_figures}
\def \tablefolder{tables}

\def \RSink {R_{sink}}
\def \RDisk {R_d}
\def \MDisk {M_d}
\def \MStar {M_*}
\def \MSol {M_{\sun}}
\def \AU {~\mathrm{AU}}

\def \Qeff {Q_{eff}}

\title[Backus \& Quinn PPD fragmentation]{Fragmentation of protoplanetary disks around M-dwarfs}

\author[Backus \& Quinn]{
	Isaac Backus,$^{1}$\thanks{E-mail: ibackus@gmail.com}
	Thomas Quinn,$^{1}$\thanks{E-mail: trq@astro.washington.edu}
	\\
	$^{1}$University of Washington, Seattle, WA 98195, USA\\
}

\date{Accepted XXX. Received YYY; in original form ZZZ}

\pubyear{2016}

\begin{document}
	\label{firstpage}
	\pagerange{\pageref{firstpage}--\pageref{lastpage}}
	\maketitle

\begin{abstract}
 
We investigate the conditions required for planet formation via gravitational
instability (GI) and protoplanetary disk (PPD) fragmentation around M-dwarfs.  Using a suite of 64 SPH simulations with $10^6$ particles, the parameter space of disk mass, temperature, and radius is explored, bracketing reasonable values based on theory and observation.  Our model consists of an equilibrium, gaseous, and locally isothermal disk orbiting a central star of mass $\MStar=\MSol/3$.  Disks with a minimum Toomre $Q$ of $Q_{min} \lesssim 0.9$ will fragment and form gravitationally bound clumps. 
\update{
Some previous literature has found $Q_{min} < 1.3-1.5$ to be sufficient for fragmentation.
Increasing disk height tends to stabilize disks, and when incorporated into $Q$ as $\Qeff\propto Q(H/R)^\alpha$ for $\alpha=0.18$ is sufficient to predict fragmentation.   
}
Some discrepancies in the literature regarding $Q_{crit}$ may be due to different methods of generating initial conditions (ICs). A series of 15 simulations demonstrates that perturbing ICs slightly out of equilibrium can cause disks to fragment for higher $Q$.  Our method for generating ICs is presented in detail.  We argue that GI likely plays a role
in PPDs around M-dwarfs and that disk fragmentation at large radii is a
plausible outcome for these disks.
     
\end{abstract}

\begin{keywords}
	accretion, accretion disks -- protoplanetary disks -- methods: numerical
\end{keywords}

\section{Introduction}
\label{sec:Introduction}

The importance of gravitational instabilities (GI) in the evolution of
protoplanetary disks (PPDs) and in planet formation remains hotly debated
\citep{cameron1978,boss1997,durisen2007,boley2010,paardekooper2012}.  In recent years, the core accretion (CA) plus gas
capture model of giant planet formation has received much attention
\citep{pollack1996,ppvi-helled}, but GI is still seen as a candidate for the
direct formation of giant planets, especially at large orbital radii
\citep{boley2009}.  While CA gives a more natural explanation of terrestrial planet formation and small bodies, GI may be important for the formation of these objects via 
solid enhancement within spiral arms or fragments \citep{haghighipour2003}.  
\update{
GI may also play an important role during the embedded phase of 
star formation \citep{vorobyov2011}.
}

Understanding the role of GI in planet formation will require continued observation of PPDs \citep{andrews2005,isella2009,mann2015} and further theoretical work.  Of primary importance are (i) disk cooling times
\citep{gammie2001,rafikov2007,meru2011,meru2012}, which must be sufficiently
short to allow density perturbations to grow against pressure support, and (ii) the
Toomre $Q$ parameter \citep{toomre1964}: 
\begin{equation} \label{eq:ToomreQ} Q
\equiv \frac{c_s \kappa}{\pi G \Sigma} 
\end{equation} 
where $c_s=\sqrt{\gamma k_B T/m}$ is the gas sound speed, $\kappa$ is the epicyclic frequency ($\kappa=\Omega$ for a massless disk), and
$\Sigma$ is the disk surface density.  As $Q$ decreases toward unity, PPDs
become increasingly unstable, and if $Q$ becomes sufficiently small, disks will
undergo fragmentation.

The parameters required for fragmentation, such as disk mass ($\MDisk$),
disk radius ($\RDisk$), and disk temperature ($T$), are constrained by the
critical $Q$ required for fragmentation.  Some previous studies have found
values of $Q_{crit} = 1.3-1.5$ \citep{boss1998,boss2002,mqws}, although it has
been noted that $Q$ can drop below unity and the disk may still tend to a
self-regulating state \citep{boley2009}.

\update{
Determining parameters required for fragmentation is complicated by issues of
resolution.  The constant ($\beta$) cooling simulations of \cite{meru2011}
demonstrated non-convergence of SPH simulations.  Further work
\citep{meru2012,rice2012} suggested artificial viscosity is to
blame. Work is underway to investigate this problem; however, resolution dependent effects are still poorly understood in SPH simulations of PPDs \citep{rice2014}.
}

Previous work has tended to focus on PPDs around solar mass stars.  Motivated by
the large population of low mass stars, we
study GI around M-dwarfs with mass $\MStar=\MSol/3$.  Around 10\% of known
exoplanets are around M-dwarfs \citep{exoplanets.org}. Due to selection
effects of current surveys such as Kepler \citep{borucki2010}, this is expected to be
a large underestimate of the actual population.  Recent discoveries show that
disks around M and brown dwarfs are different from those around solar analogs:
the mass distribution falls of more slowly with radius, and is denser at the
midplane. These differences change disk chemistry and the condensation sequence.
M-dwarf disks are also less massive and survive longer
\citep{apai2009,apai2010}.  Core accretion timescales, which scale as the orbital period, are long around M-dwarfs.  Because the stars are much lower in luminosity, their
disks are substantially cooler.  Additionally, planets orbiting nearby M-dwarfs
are likely to be the first smaller planets spectroscopically characterized
\citep{seager2015}.

\update{
The simulations of \cite{boss2006a} indicate that GI is able to form gas giants
around M-dwarfs.  \cite{boss2006b} and \cite{boss2008} even argue that super 
earths around M-dwarfs can be explained as gas giants, formed via GI, and stripped of their gaseous envelopes by photoevaporation.
}

In this paper we explore the conditions required for disk fragmentation under GI
around M-dwarfs.  Previous studies have found a range of values of the
$Q_{crit}$ required for disk fragmentation
\citep{boss1998,boss2002,mayer2008,boley2009}.  Discrepancies may be due to
different equations of state (EOS), cooling algorithms, numerical issues such as
artificial viscosity prescriptions, Eulerian vs. Lagrangian codes, and initial
conditions (ICs).  ICs close to equilibrium are non-trivial to produce and so we explore the dependence on ICs of simulations of
gravitationally unstable disks.

\update{
We focus on probing disk fragmentation around M-Dwarfs, which remains poorly studied, and the importance of ICs.  These warrant a simple, well understood isothermal EOS.  We therefore probe the Toomre $Q$ required for fragmentation 
while leaving the question of the cooling required for fragmentation for future work.
}

We begin in \S\ref{sec:ICgen} by presenting our method for generating
equilibrium initial conditions for smoothed-particle hydrodynamic (SPH)
simulations of PPDs, with particular care taken in calculating density and
velocity profiles.  \S\ref{sec:SetOfRuns} describes the suite of
simulations presented here and discuss the theoretical and observational
motivations behind our disk profiles.  \S\ref{sec:FragmentationAnalysis}
presents our analysis of disk fragmentation around M-dwarfs and discusses the
importance of ICs in simulations of unstable disks.   \S\ref{sec:Clumps} 
presents our method for finding and tracking gravitationally bound clumps and
discusses clump formation in our simulations.  We present our discussion in
\S\ref{sec:Discussion}.  We consider the effects of thermodynamics and ICs on
fragmentation in PPD simulations and argue that GI should play an important role
in PPDs around M-dwarfs and that we expect disk fragmentation at large radii to
occur around many M-dwarfs.

\section{Initial Conditions}%
\label{sec:ICgen}

A major emphasis of this work was to ensure that ICs were as close to
equilibrium as possible.  Axisymmetric disks very near equilibrium may not realistically model actual PPDs, but we wish to make as few assumptions as possible about disks and have attempted to minimize numerical artifacts.  One worry is that disks too far from equilibrium may
artificially enter the non-linear regime, possibly initiating fragmentation in
an otherwise stable disk.  This possibility is explored in
\S\ref{sec:SensitivityToICs}.  In this section, we present our method for generating ICs.\footnote{Our code for generating ICs is freely available on github at \url{https://github.com/ibackus/diskpy} as a part of our PPD python package \textit{diskpy}}

Many methods have been used in previous work for generating initial conditions. 
As discussed by \cite{mayer2008}, apparently contradictory results in
fragmentation studies may be due to differences in ICs used.  Much published
research does not detail IC generation in sufficient detail to be reproducible,
but we can sketch out a few different approaches used.  \cite{boss1998}
developed ICs by defining the midplane density $\rho(R, z=0)$, analytically
estimating $\rho(z)$, using an approximate circular velocity, and iteratively
adjusting the temperature profile to create a steady state solution.

As with us, other authors \citep{mqws,rogers2011} were able to to define the surface density and temperature profiles.  They then estimated vertical hydrostatic equilibrium to
calculate density.  \cite{mqws} estimated the gas velocity required for circular orbits ($v_{circ}$) from gravitational forces and
adjusted for hydrodynamic forces.  Additionally, as with others (e.g.
\cite{mayer2008}), they also approached low $Q$ values by slowly growing the
disk mass.  Similarly, \cite{boley2009} used low mass, high-$Q$ models in his grid code simulations and
accreted mass from the $z$ boundaries gradually to grow simulations towards
instability.

\cite{pickett2003} placed great care in developing equilibrium ICs.  In contrast
to our ICs, they also modeled the central, accreting star.  They generated a
stable disk ($Q=1.8$) using a field equilibrium code \citep{hachisu1986}.  They
specified the specific angular momentum $j(R)$ of the gas (which forces velocity to be solely a
function of radius) then iteratively used a self consistent field method to
solve the Poisson gravity equation and balance the hydrodynamic forces to
approach equilibrium.  A shooting method for $j(R)$ was used to reach a desired
$\Sigma(R)$.  For low-$Q$ simulations, they cooled the disk until it reached
$Q_{min}=0.9$.

For our simulations, we desired to scan parameter space by defining surface
density ($ \Sigma $) and temperature ($ T $) radial profiles, along with star
mass.  From these, the gas density ($ \rho $) can be estimated to ensure vertical
hydrostatic equilibrium in the disk.  SPH particles are then semi-randomly
seeded and their equilibrium circular velocities are estimated using the NBody/SPH simulation code ChaNGa to
calculate the forces.  Our method allows us to directly and quickly generate equilibrium ICs for low $Q$ values and arbitrary $\Sigma$ and $T$ profiles.

\subsection{Estimating $ \rho(R,z) $}%
\label{sec:RhoEst}

To estimate $\rho(R,z)$, we first define $\MStar$, $\Sigma(R)$, and $T(R)$.
Hydrostatic equilibrium is solved by adjusting the vertical density profile to
maintain vertical hydrostatic equilibrium and adjusting the gas orbital velocity
to ensure radial equilibrium.

To be in equilibrium along the vertical direction, the vertical component of
gravity from the star and the disk's self-gravity should balance the vertical
pressure gradient in the gas.  For the disk self-gravity term, we assume the
thin disk approximation where $ \rho $ is assumed to be only a function of $ z
$. $ T $ is set to be independent of $ z $, which is reasonable for the locally isothermal equation of state used in these simulations. 
All quantities are axisymmetric 
and symmetric about the midplane $ z=0 $. Under these assumptions, the vertical hydrostatic
equilibrium condition can be written as:
\begin{equation}
    \label{eq:Equilibrium} \frac{k_BT}{m} \frac{d\rho}{dz} + \frac{G M_*
    z \rho}{(z^2+R^2)^{3/2}} + 4\pi G\rho \int_0^z \rho(z')\,dz'=0
\end{equation}
where $ m $ is the mean molecular weight of the gas, $ M_* $ is the
star's mass and $ R $ is the cylindrical radius.  The first term is the pressure gradient and would be altered for a non-isothermal EOS.  The boundary
conditions are:
\begin{subequations}
    \label{eq:BCs}
    \begin{equation}
        \label{eq:BC1} \int_{0}^{\infty}\rho(z')dz' = \Sigma/2
    \end{equation}
    \begin{equation}
        \label{eq:BC2} \frac{d\rho}{dz}\Bigr|_{z=0}=0
    \end{equation}
\end{subequations}

To apply these boundary conditions, eq.(%
\ref{eq:Equilibrium}) is transformed to be a differential equation for $
I(z) \equiv\int_{z}^{\infty}\rho(z')\,dz' $, which gives an equation of
the form:
\begin{equation}
    \label{eq:EqI} \frac{d^2I}{dz^2} + \frac{dI}{dz}\left[\frac{c_1z}{(z^2+R^2)^
    {3/2}} + c_2\left(\frac{\Sigma}{2}-I\right)\right]=0
\end{equation}
with the constants set by eq.(%
\ref{eq:Equilibrium}).  Then eq.(%
\ref{eq:EqI}) is solved numerically by a root finding algorithm.  The
density is calculated by applying $ \rho=-\frac{dI}{dz} $.  To ensure a
robust solution, the solution for $ \rho $ is then iteratively entered
into a root finding algorithm for eq.(%
\ref{eq:Equilibrium}) and rescaled to ensure the boundary condition in
eq.(%
\ref{eq:BC1}) is met.
\footnote{The current publicly available version of our code \textit{diskpy} uses an altered version of this algorithm to solve eq.~\ref{eq:Equilibrium} which is significantly faster and has been tested to provide the same results as the algorithm described here.}

\subsection{Generating particle positions}%
\label{sec:PosGen} The solution to $ \rho $ is then used to
semi-randomly seed SPH particles.  To mitigate Poisson noise, we used the method of \cite{csw2009} which places the particles along a spiral in the $ x-y $ plane to keep them more evenly spaced.  The spiral is made in such a way that $\Sigma(R)$ is reproduced. See \cite{csw2009} for a more detailed explanation.  The particle $ z $ values are then randomly assigned
according to $ \rho(z) $.  
Note that it is important to make the spirals trailing (rather than leading)
to avoid unwinding effects and swing amplification.

\subsection{Circular velocity calculation}%
\label{sec:VelocityCalculation}

A major concern with avoiding artifacts in the ICs is calculating the particle
velocities required for circular orbits ($ v_{circ} $).  For typical disks, $
v_{circ} $ is one to two orders of magnitude larger than $ c_s $.  Thus,
particle velocities which deviate from $ v_{circ} $ on the percent level will
deposit a significant amount of power into the disk.  If this dominates the
available thermal energy (which tends to stabilize the disk), the disk may enter
a non-linear regime and fragment for unrealistically high $ Q $.  We include a
demonstration of this in \S\ref{sec:SensitivityToICs}.

To calculate $ v_{circ} $, we employ our simulation code ChaNGa 
to calculate the radial gravitational and hydrodynamical forces on all
particles. Following \cite{mqws}, we set the gravitational softening length to $
\epsilon_s=0.5 \expectation{h} $, where $ \expectation{h} $ is the SPH smoothing
length, calculated over the 32 nearest neighbors, and averaged over all
particles in the simulation.  The softening length for the star is set as the
distance to the nearest gas particle.  ChaNGa is then used to estimate the
gravitational and SPH forces separately.

To deal with SPH noise, the radial forces must be averaged over many particles. 
The gravitational and SPH forces are averaged separately because of a different
dependence on position.  In both cases, 50 radial bins are used.  To fit the
 $R$ and $z$ dependence of the gravitational forces, particles are binned
radially, and for each bin a line is fit to the radial force per mass due to
gravity, as a function of $ \cos\theta $:
\begin{equation}
    a_{g,i}(\cos\theta)=m_i\cos\theta+b_i
\end{equation}
where $ \theta $ is the angle above the plane of the disk.  Here $ \cos\theta
$ is chosen as the independent variable because the radial gravitational
force from the central star is proportional to $ \cos\theta $.  The fit
parameters $ (m_i, b_i) $ are then linearly interpolated as a function
of $ R $ to calculate $ a_g(R,z) $.

The radial hydrodynamic forces display very little $ z $ dependence, as
is expected for a temperature profile independent of $ z $.  The SPH
forces are averaged over logarithmically spaced radial bins and then
interpolated with a linear spline.  From the total force per mass ($ a=a_
{grav}+a_{SPH} $) we can calculate $ v_{circ} $ as:
\begin{equation}
    v_{circ} = \sqrt{a R}
\end{equation}

\section{Set of runs}%
\label{sec:SetOfRuns}

\subsection{ChaNGa}%
\label{sec:ChaNGa}
All the simulations in this paper were performed with ChaNGa\footnote{A public version of ChaNGa can be downloaded from \url{http://www-hpcc.astro.washington.edu/tools/changa.html}}, a highly
parallel N-body/SPH code originally written for cosmology simulations
\citep{Menon15}.  ChaNGa is written in the {\sc Charm++} parallel
programming language which allows the overlap of computation and
communication, and enables scaling to over half a million processor cores
\citep{Menon15}.  The physics modules of ChaNGa are taken from {\sc
  gasoline} \citep{Wadsley04} where gravity is calculated using a
Barnes-Hut \citep{barnes86} tree with hexadecapole expansions of the
moments, and hydrodynamics is performed with SPH.  All the simulations
performed here used a gravitational force accuracy (node opening)
criterion of $\theta_{BH} = 0.7$.  Timesteps are set by a Courant
condition of $\eta_C = 0.3$ and an acceleration criterion of $\Delta
t_i = \eta \sqrt{\epsilon_i \over a_i}$ where $\epsilon_i$ and $a_i$ are
respectively the softening and acceleration of a particle, and $\eta
= 0.2$.  To stabilize the SPH we use the \cite{monaghan1992} artificial
viscosity
with coefficients $\alpha = 1$ and $\beta = 2$, and we use the
\cite{balsara95} switch to suppress viscosity in non-shocking
shearing environments.  We have tested ChaNGa with the PPD Wengen code
test 4 and it performs well (see Appendix \ref{appendix:Wengen}).  The simulations presented here each took about 3k core-hours running on eight 12-core nodes on the University of Washington's supercomputer Hyak.

\subsection{Disk Profiles}%
\label{sec:Profiles}

\begin{figure}
	\includegraphics[width=\columnwidth]{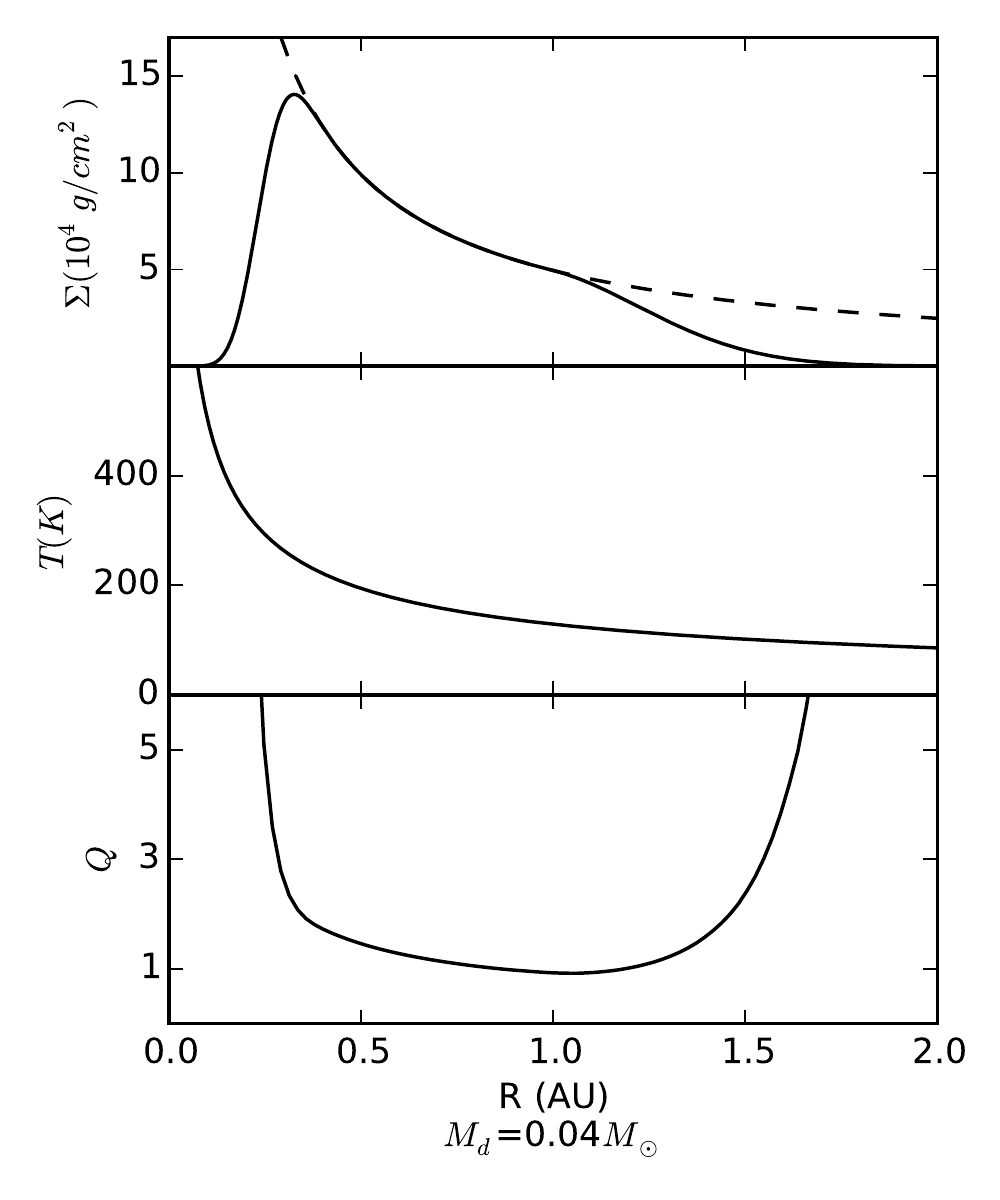}
	\caption{ Example radial profiles for powerlaw surface density $
		\Sigma\propto 1/R $, with $ R_{in}=0.3 \AU $ and $ \RDisk=1\AU $.
		\textbf{Top:} Surface density profile including cutoffs (solid) and
		excluding cutoffs (dashed). \textbf{Middle:} Disk temperature $ T\propto
		R^{-0.59} $ \textbf{Bottom:} Toomre $ Q $, calculated including full disk self gravity, SPH forces, and calculation of $\kappa$.%
		\label{fig:ExProfPowerlaw}}
	
\end{figure}

\begin{figure}
	 \includegraphics[width=\columnwidth]{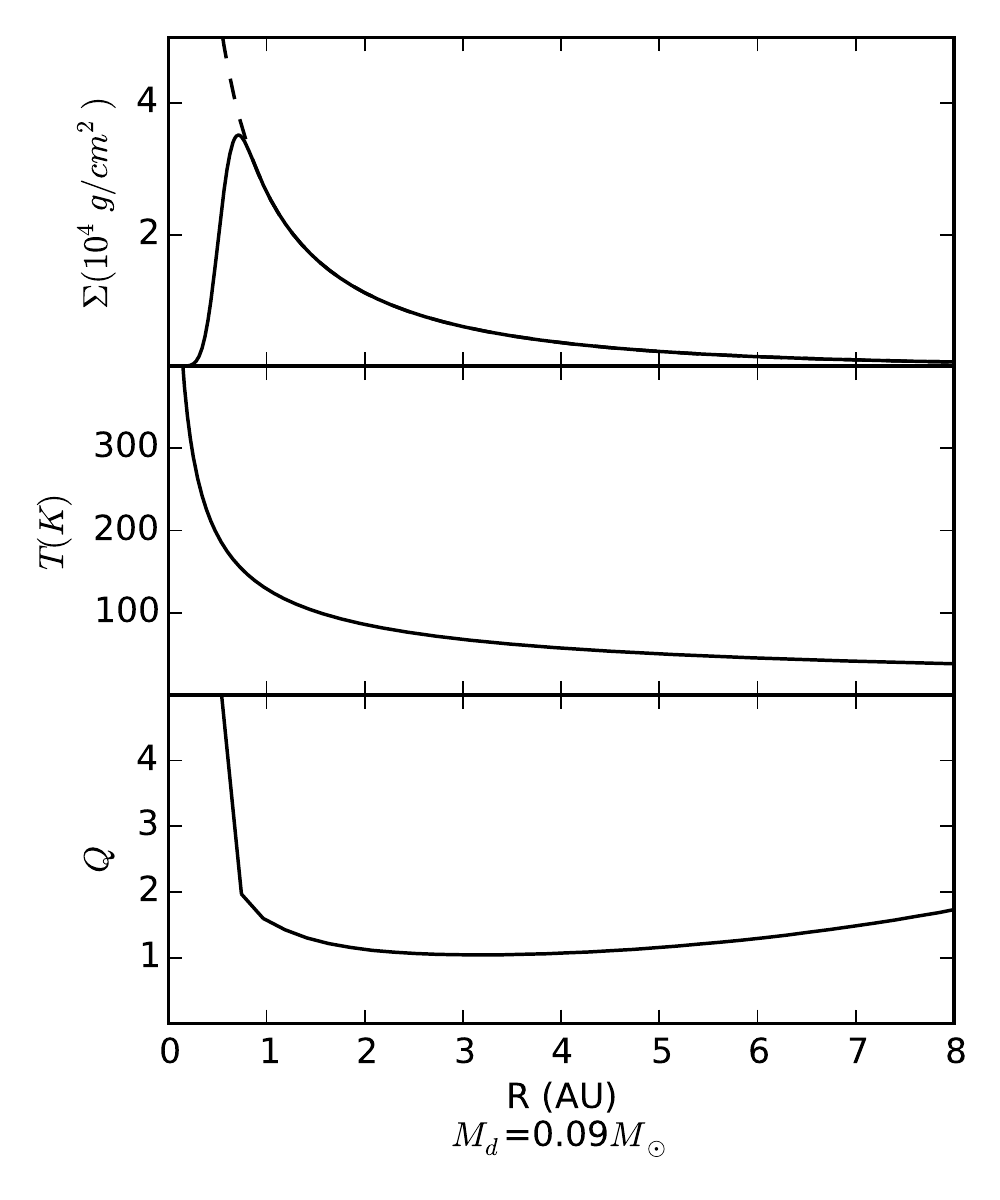}
	\caption{ Example radial profiles for a viscous disk surface density
		$ \Sigma(r)=\Sigma_0 r^{-\gamma} \exp(-r^{2-\gamma}) $, where $ r $
		is a dimensionless radius and $ \gamma=0.9 $.  The radius containing
		95\% of the mass is $ \RDisk=11\AU $. \textbf{Top:} Surface density
		profile including cutoffs (solid) and excluding cutoffs (dashed).
		\textbf{Middle:} Disk temperature $ T\propto R^{-0.59} $ \textbf{Bottom:}
		Toomre $ Q $.%
		\label{fig:ExProfViscous}}
	
\end{figure}

Figures~%
\ref{fig:ExProfPowerlaw}~\&~%
\ref{fig:ExProfViscous} show radial profiles for example simulations.
Pictured are $ \Sigma(R) $ (top panel), $ T(R) $ (middle panel), and
$ Q(R) $ (bottom panel).  The exact disk structure of PPDs, especially young
ones, is poorly constrained.  Therefore we
adopt simple, easy to interpret profiles, and consider a range of values of
$\MDisk$ and $T$ in order to bracket plausible disk parameters.  Below we
describe our choice of disk temperature and surface density profiles, along with
the theoretical and observational motivations for them.

\subsubsection{Temperature}%
\label{sec:TemperatureProfiles} For a blackbody disk with heating
dominated by solar radiation,
\cite{chianggoldreich1997} showed that the temperature profile should be
a power law.  The exponent is $ {q=3/7} $ for a fully flared disk and $ {q=3/4}
$ for a flat disk.  We therefore adopt a temperature profile of the
form:

\begin{equation}
    \label{eq:T} T(R)=T_0\left(\frac{R}{R_0}\right)^{-q}
\end{equation}
where $ {R_0} $ was set to be 1~AU and $ {T_0} $ is the
temperature at 1~AU.  The values we adopt for $ T_0 $ and $ q $ come
from the observations of
\cite{andrews2005}.  They observed dust SEDs of circumstellar disks in
the Taurus-Auriga star forming region.  Fit results for 44 mainly solar
type stars yielded median inferred parameters of $ {T_0=148K} $ and $ {q=0.58} $
(see their Table~2).  Averaging their results for M-Stars only, we adopt
$ {q=0.59} $ for every simulation and $ {T_0=130K} $ as our central
fiducial values.  This power law of $ {q=0.59} $ lies between a fully
flared and completely flat disk.  Given the uncertainty of these values
we also ran simulations with $ {T_0=65K} $ and $ 260K $ to bracket plausible disk temperatures.

\subsubsection{Surface density}%
\label{sec:SurfaceDensityProfiles}

Two functional forms for $ \Sigma $ were used:  a power law (Fig.~%
\ref{fig:ExProfPowerlaw}) and the similarity solution for a thin,
viscous disk (Fig.~%
\ref{fig:ExProfViscous}).  The power law used was:
\begin{equation}
    \label{eq:SigmaPowerlaw} \Sigma(R)=\Sigma_0 \left(\frac{R}{R_0}\right)^
    {-1}
\end{equation}
Where the normalization $ \Sigma_0 $ is fixed by the desired disk mass.
As shown in Fig.~%
\ref{fig:ExProfPowerlaw}, interior and exterior cutoffs were applied.
For $ R>\RDisk $, an exponential cutoff was applied by multiplying $
\Sigma(R) $ in eq.(%
\ref{eq:SigmaPowerlaw}) by:
\begin{equation}
    \label{eq:ExpCutoff} \Sigma_{exterior}=\Sigma(R) \mathrm{e}^{-(R-\RDisk)^2/L^2}
\end{equation}
where the cutoff length was set to $ L=0.3\RDisk $.  
This form ensures that $\Sigma$ and $\frac{d\Sigma}{dR}$ are unchanged at $R=R_d$. The interior cutoff 
was applied by multiplying $ \Sigma $ by a smooth high order polynomial
approximation to a step function, defined to be $ [0,1] $ at $ R=[0,R_{cut}]
$ with the first 10 derivatives set to be 0 at $ R=[0,R_{cut}] $.
For these simulations $ R_{cut}=0.5 \RDisk $ and $ \Sigma $ differs
significantly from a power law for $ R\lesssim0.3 \RDisk $ (see Fig.%
\ref{fig:ExProfPowerlaw}).  This radius was chosen such that:  (a) $ Q\gg1
$ at $ R_{cut} $ to ensure the disk is stable at $ R_{cut} $, and (b)
the cutoff is applied far enough from the most unstable disk region and
removes little enough mass that fragmentation should not be strongly
affected by the cutoff.

The second functional form for $ \Sigma $ (see Fig.%
\ref{fig:ExProfViscous}) comes from the similarity solution to a thin,
light, viscous disk orbiting a star, as found in
\cite{bellpringle1974}.  For a viscosity obeying a power law $ \nu\propto
R^\gamma $, at a given time $ \Sigma $ can be written as:
\begin{equation}
    \label{eq:SigmaViscous} \Sigma(r)=\Sigma_0 r^{-\gamma} \exp(-r^{2-\gamma})
\end{equation}
where $ r $ is a dimensionless radius and the normalization $ \Sigma_0 $
is fixed by the disk mass.  Note that the full similarity solution
includes a time dependence which we fold into $ \Sigma_0 $ and $ r $.
From fits to observations of 9 circumstellar disks,
\cite{andrews2009} found a median value of $ \gamma=0.9 $, which is the
value we adopt here.  For this profile, no exterior cutoff is required.
The same interior cutoff as for the power law $ \Sigma $ was applied at $
r=0.1 $.

If we define $ \RDisk $ to be the radius containing 95\% of the disk
mass (ignoring the interior cutoff), then:
\begin{equation}
    \label{eq:r} r=\frac{R}{\RDisk} \ln(1/0.05)^{1/(2-\gamma)}
\end{equation}

As has been noted before (e.g. \citet{isella2009}), there is no physical
motivation for adopting a power law for $ \Sigma $.  It is also not clear how applicable the viscous profile is.  These are just simple functional forms often adopted in previous work.  We have examined the end state of high-$Q$ runs and they are better approximated by the viscous profile, but the fit is not perfect.

\subsection{Run Parameters}%
\label{sec:RunParameters}

For a given functional form of $ \Sigma(R) $ and $ T(R) $, three
parameters must be set to make our ICs and thereby fix a value of $ Q_{min}
$:  the temperature normalization ($ T_0 $), the disk mass ($ \MDisk $),
and the disk radius ($R_d$).  The choices of $ T_0 $ are discussed in \S\ref{sec:TemperatureProfiles}.
\subsubsection{Disk mass}%
\label{sec:DiskMass}

Under our scheme, setting $ \MDisk $ fixes the surface density normalization. 
We selected plausible values to explore, ranging from $ 0.01 $ to $ 0.08 \MSol
$. \cite{isella2009} reported observations of 11 disks around pre-main-sequence stars,
including 7 around M-stars.  From their Tables 1 \& 2, we find a median value of
$ \MDisk/\MStar=0.15 $, which for our simulations ($ \MStar=1/3 \MSol $) gives a
central value of $ \MDisk=0.05\MSol $.  It should be stressed that these disk masses are inferred using an assumed gas to dust ratio of 100 and therefore may have large, uncharacterized uncertainties.

\subsubsection{Disk radius}%
\label{sec:DiskRadius}

\cite{isella2009} argue that for a typical disk, $ R_d $ increases from
around $ 20\AU $ to $ 100\AU $ over the course of $ \sim $5~Myr.  This
tends to stabilize disks by decreasing $ \Sigma $ over time.  Since we
are interested in disks at their most unstable, we adopt $ 20\AU $ as
our central fiducial value.  To explore parameter space, we used values
of $ \RDisk $ ranging from $ 1/3\AU $ to $ 30\AU $.

\subsubsection{Numerical Parameters}%
\label{sec:NumericalParameters}

For the analysis of fragmentation criteria, a set of 64 ICs were run
with $ 10^6 $ particles.  For typical disks, this yields particles with
a mass of $ m_{particle}\approx 5\scinote{-5} M_{Jupiter} $.  A locally
isothermal EOS with a mean molecular weight of 2 was used for the gas.
Following
\cite{mqws}, we set the gravitational softening length to $ \epsilon_s=0.5
\expectation{h} $, where $ \expectation{h} $ is the SPH smoothing length
calculated over the 32 nearest neighbors and averaged over all
particles.  Typical values are around $ {\expectation{h}=5\scinote{-3}
\RDisk} $.

The central star (mass $ \MSol/3 $) was set to be a sink particle:
when a gas particle approaches the star within a distance of $ \RSink $,
its mass and momentum are accreted onto the star.  $ \RSink $ was set as
the distance to the closest gas particle in the ICs.  This yielded $
\RSink=[0.08 \RDisk, 0.02 \RDisk] $ for the power law and viscous $
\Sigma $ profiles, respectively

Making the central star a sink serves two purposes worth mentioning.  A
gas particle near the star gains a large velocity, experiences strong
forces, and is often captured in a tight orbit around the star.  This
requires a very small time-step which can increase computation time by
orders of magnitude.  Secondly, the adaptive time-stepping used by
ChaNGa can fail to conserve momentum when two interacting particles require
time-steps which differ by more than a couple orders of magnitude.  This
momentum non-conservation is particularly problematic for tight, rapid
orbits.  The effect is sufficiently strong that very stable disks ($ Q>2 $)
were seen to fragment when the central star was not treated as a sink.

We note that treating the star as a sink in this manner is unrealistic in that
accretion happens at very large radii (of order an$\AU$).  When accretion
happens, the star will jump to the center of mass of the star + particle system.
In the limit of low accretion rates and in locations far from the star this
will be a negligible effect, but in disks with high accretion a different 
scheme should be used.  For our simulations there is very little
accretion: less than $10^{-3}\MDisk$ over the duration of the simulations.

As shown in figure~%
\ref{fig:Runtime}, non-fragmenting simulations were run for $ \sim $30
outer rotational periods (ORP), where we define 1~ORP as the orbital period
at the most unstable disk radius.  ORPs for our disks range from 0.3~yrs for our smallest disks to 300~yrs for the largest.  Disks with parameters close to
fragmentation were run longer (100--200~ORP) to ensure they reached a
steady state which would not fragment.  Since the main goal of this work
is to investigate fragmentation of PPDs and since computation time
increases drastically after clump formation, simulations which
fragmented were run for around 1--2~ORP after fragmentation.  Figure~%
\ref{fig:Runtime} shows the fragmentation timescales for these disks.

\section{Fragmentation analysis}%
\label{sec:FragmentationAnalysis}

\begin{figure}
	 \includegraphics[width=\columnwidth]{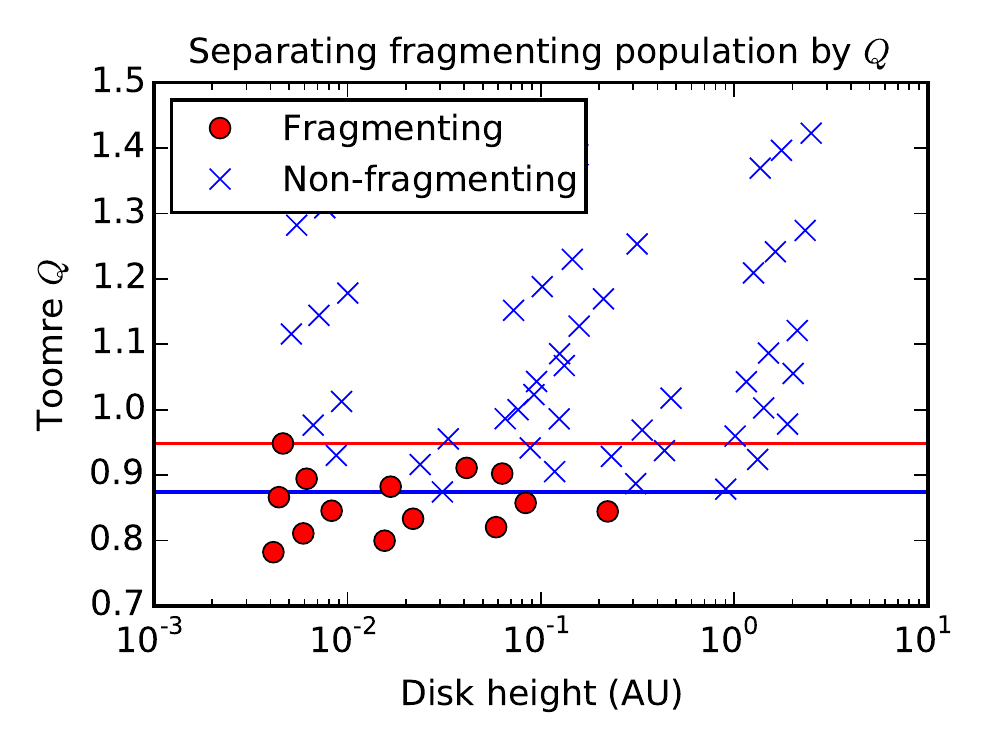}
	\caption{Minimum Toomre $ Q $ for the fragmenting (clump--forming)
		and non-fragmenting simulations.  The red (blue) lines mark the largest (smallest) values of the fragmenting (non-fragmenting) simulations.  The two populations overlap around
		$ Q\approx0.9 $.%
		\label{fig:QFrag}}
\end{figure}

\begin{figure}
	 \includegraphics[width=\columnwidth]{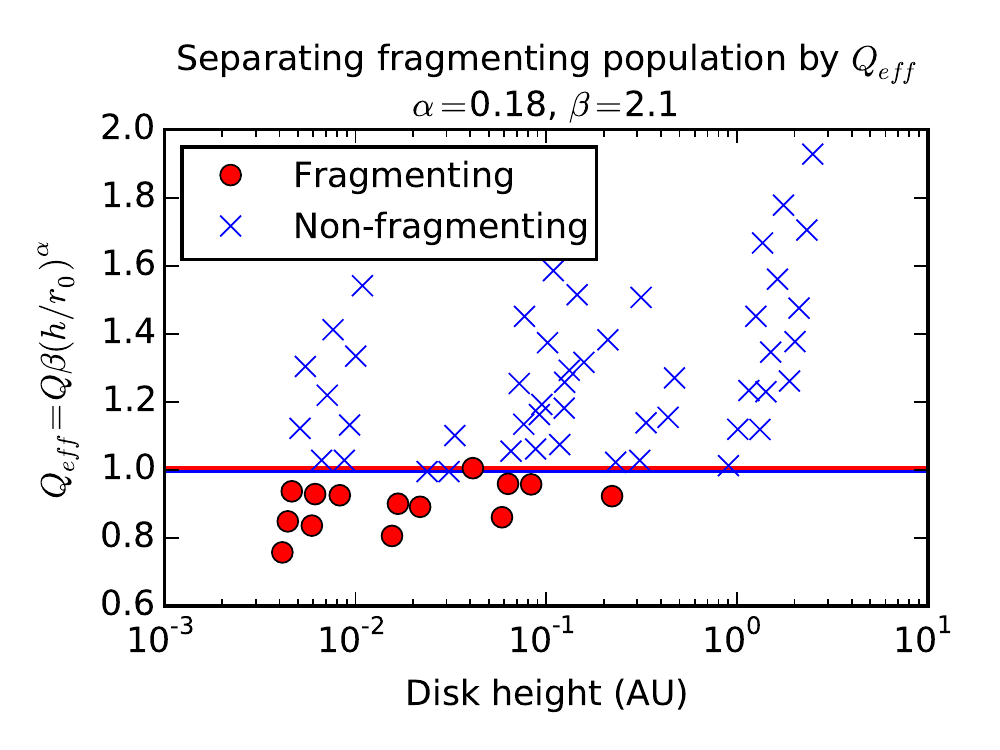}
	\caption{Minimum effective Toomre $ Q $ ($ \Qeff $) for the
		fragmenting (clump--forming) and non-fragmenting simulations.
		Re--parameterizing the stability criterion for a protoplanetary disk
		as $ \Qeff=Q\beta(H/R)^\alpha $ is sufficient for predicting
		whether a protoplanetary disk will fragment.  $\beta$ is a normalization
		factor chosen such that disks with a $ \Qeff < 1 $ will fragment.%
		\label{fig:QeffFrag}}
\end{figure}

The primary goal of this paper is to investigate under what conditions
we can expect a PPD surrounding an M-dwarf to fragment.  The parameters
explored by our model are $ T $, $ \MDisk $, and $ \Sigma $.  Although
all simulations were run with a star of $ \MStar=\MSol/3 $, we also
extend our analysis to stars of similar mass.  As expected, we find sufficiently heavy or cold disks will fragment under GI.

Gravitational instability in PPDs is typically parameterized by the
Toomre $ Q $ parameter (eq.~\ref{eq:ToomreQ}).  For our isothermal
simulations, $ c_s=\sqrt{k_B T/m} $.  Two dimensional disks will be unstable for $ Q>1 $, but the instability and
fragmentation criteria for 3D disks remain uncertain.  Previous studies
have found that 3D disks will fragment for $ Q_{min} $ significantly
greater than 1 \citep{boss1998,boss2002,mqws,durisen2007}; however, we do
not find this to be the case for our simulations.  Figure~%
\ref{fig:QFrag} shows the fragmentation boundary for our simulations.
Disks with $ Q\lesssim0.9 $ fragment.  Here, we define fragmentation to be when the first gravitationally bound clump is found (see~\S\ref{sec:Clumps}).

\update{
We also considered the effects of swing amplification.  If the parameter:
\begin{equation}
X_m = \frac{\kappa^2 R}{2\pi G \Sigma} \frac{1}{m}
\end{equation}
is near or below unity, small leading disturbances will amplify upon becoming 
trailing disturbances and can drive disk dynamics, especially spiral arm growth
\citep{binneytremaine1987}.  Lower order modes are expected to dominate. 
For our disks, our lowest value is $X_1 = 5.7$, and most disks have $X_1 > 10$,
so swing amplification should not be significant in these simulations.
}

\secondupdate{Another proposed instability in gaseous disks is provided by the SLING mechanism \citep{adams1989,shu1990}, which is driven by $m=1$ mode growth and is sensitive to the outer edge of the disk.  For our simulations, we see little $m=1$ power.  Additionally, we tested the importance of the outer edge by applying a step-function cutoff to $\Sigma$ on the disk outer edge for marginally stable disks and the overall behavior was unaltered.  It therefore seems unlikely that the SLING mechanism plays a major role in these simulations.}

It should be noted that for our simulations we calculate $Q$ from the ICs. 
\update{
Since the equilibrium orbital velocity is known (see
\S\ref{sec:VelocityCalculation}), we can directly calculate $\kappa^2 =
\frac{2\Omega}{R} \frac{d}{dR}(R^2\Omega)$.  This fully includes the effects of
a 3D disk with self-gravity and pressure gradients.  $\Sigma$ is calculated by
binning SPH particles radially, summing their masses, and dividing by the annulus
area.
}
A common approximation for light disks is to ignore disk self gravity and pressure gradients and use
$Q\approx c_s\Omega/\pi G\Sigma$, which for disks of $\MDisk/\MStar\approx 0.1$
underestimates $Q$ at the 10\% level.
\update{
Different estimates of $Q$ may account for some of the discrepancy in the
literature regarding the critcal $Q$ required for fragmentation.
}

As can be seen in figure~\ref{fig:QFrag}, there is some overlap in $Q$ for the
fragmenting and non-fragmenting populations.  Since the Toomre stability
criterion strictly applies to a 2D disk, this is unsurprising.  Following the
work of \cite{romeo2010}, to higher order a disk scale height correction enters
into the dispersion relation for axisymmetric perturbations. Taller disks should
be more stable than thinner ones.

We re-parameterized $Q$ to include disk height as:
\begin{equation}
\label{eq:Qeff}
\Qeff \equiv Q \beta (H/R)^{\alpha}
\end{equation}
where $H$ is the disk scale height and $\alpha$ is a free parameter. $\beta$ is
normalization parameter that we
set such that \mbox{$\Qeff < 1$} is the boundary for disk fragmentation (see
below).  $H$ is calculated as the standard deviation of the vertical density profile,
rather than the first order approximation $c_s/\Omega$.  We then fit
the power law $\alpha$ to
minimize the overlap of the boundaries in $\Qeff$ for the
fragmenting/non-fragmenting population.  Figure~\ref{fig:QeffFrag} shows the
separation of the two populations (compare to fig.~\ref{fig:QFrag}).  For
\mbox{$\alpha=0.18$} and \mbox{$\beta=2.1$}, simulations with \mbox{$\Qeff < 1$}
fragment.

A power law was chosen because it is a simple functional form, and there is a
great deal of self-similarity in PPDs, but other forms such as linear
corrections might be suitable as well.  As expected, taller disks have a larger
$Q_{eff}$ and are therefore less prone to fragmentation, although the dependence
on height is weak.

\begin{figure}
	\includegraphics[width=\columnwidth]{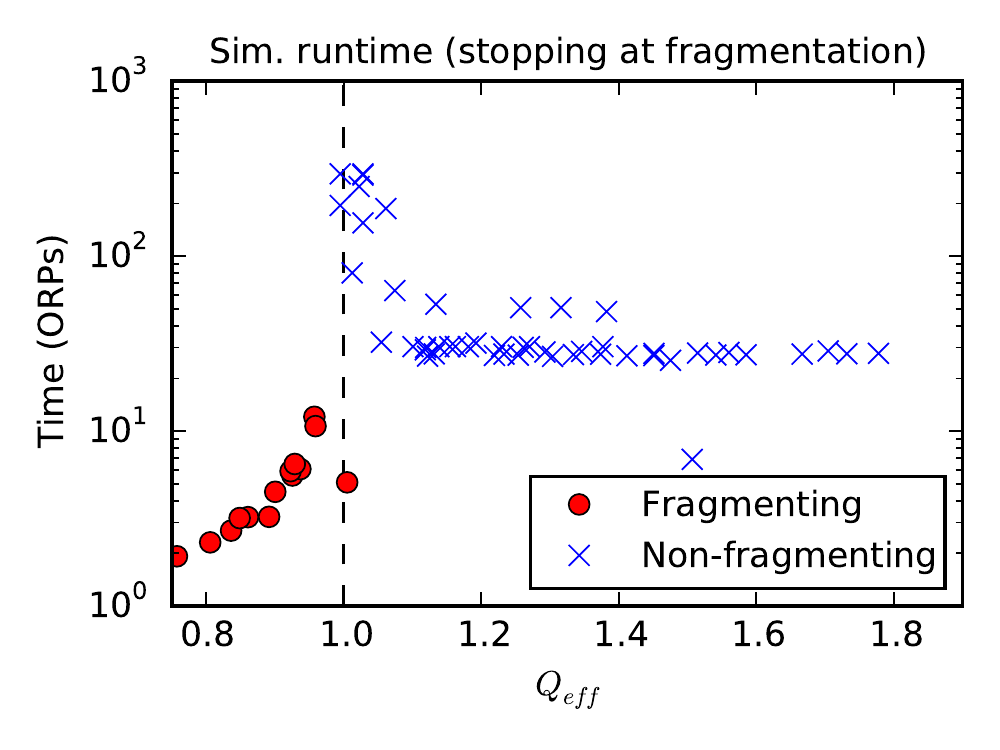}
	\caption{Total simulation time (non-fragmenting simulations) and
		time until fragmentation (fragmenting simulations), in units of the
		orbital period at the most unstable disk radius.  Fragmentation is
		defined to occur when a gravitationally bound clump forms.  The
		fragmentation timescale increases rapidly as $ \Qeff $ approaches
		1.  Simulations with $ \Qeff \gtrsim 1 $ were run for longer to
		verify that they do not fragment.%
		\label{fig:Runtime}}
\end{figure}

We also found that $\Qeff$ correlates strongly with time until fragmentation
(see fig.~\ref{fig:Runtime}).  We found it to predict fragmentation time with
less scatter than $Q$.  To verify that $H/R$ is an important parameter in
predicting disk fragmentation, we considered power law dependence of $\Qeff$ on
various dimensionless combinations of parameters, including the most unstable
wavelength, $c_s$, $T$, and $\Omega$.  We found $H/R$ to separate the
populations most strongly.

These considerations indicate disk scale height is an important parameter in
dictating stability and fragmentation.  With the fragmentation boundary
$\Qeff=1$, we are equipped to estimate disk parameters for which we may reasonably
expect disks to fragmentation.

\begin{figure}
	\includegraphics[width=\columnwidth]{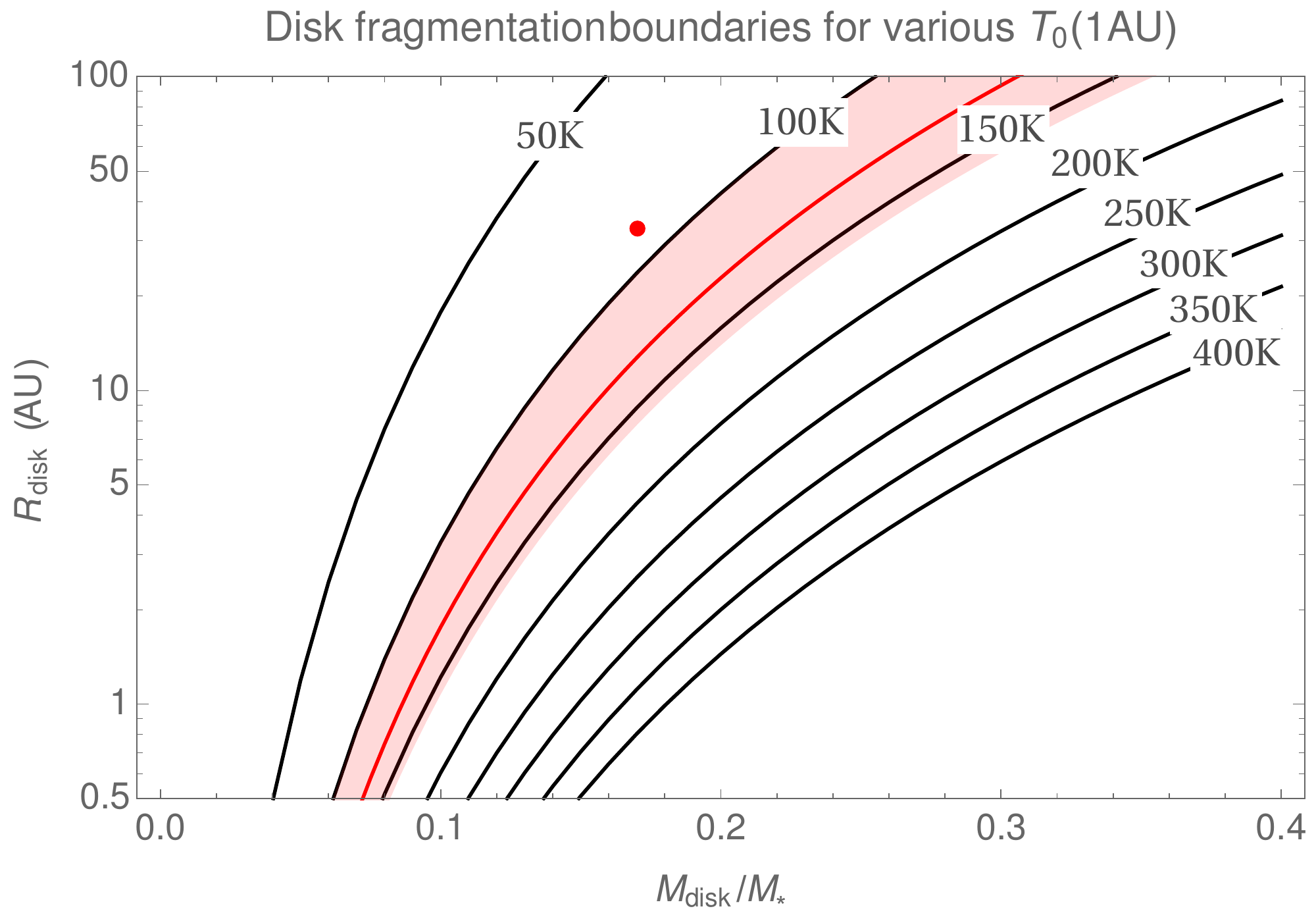}
	\caption{ Fragmentation criteria for disk ICs.  The curved lines
		define which disks will fragment for various disk temperatures at
		1~AU, assuming a temperature profile of $ T\propto r^{-0.59} $ and 
		a surface density profile of $\Sigma\propto 1/R$. 
		ICs which lie to the right of a line will fragment.  The boundaries
		are $\Qeff=1$ contours, where the $\Qeff$ estimates include
		approximations for disk height and disk self gravity.  The red line
		marks the boundary defined by the fiducial temperature of
		$T_0(1\AU)=130 \pm 25 K$ from \protect\cite{andrews2005} and the red dot
		marks the fiducial disk mass and radius for a young disk from
		\protect\cite{isella2009} (see \S\ref{sec:RunParameters} for a discussion of
		these values).  These fiducial values likely have large
		uncertainties (not pictured).%
		\label{fig:FragBound}}
\end{figure}

Figure~\ref{fig:FragBound} shows the boundaries for disk fragmentation for a
star of mass \mbox{$\MStar=\MSol/3$} as a function of $\RDisk$, $\MDisk$, and
the temperature at $1\AU$ ($T_0$).  The contour lines mark the boundary for
various values of $T_0$.  Disks to the right of the contour lines have a minimum
\mbox{$\Qeff<1$} and will fragment.

The red boundary marks the fiducial value of $T_0 = 130 K$ from the observations
of \cite{andrews2005} (see \S\ref{sec:TemperatureProfiles}), with the
surrounding red region marking the sample scatter in their observations of
$25K$.  The red point marks the fiducial values (for a young disk) of $M_d$ and
$R_d$ from the observations of \cite{isella2009}
(see~\S\ref{sec:SurfaceDensityProfiles}).

Since the fiducial disk parameters lie to the left of the fiducial boundary, we expect
the observed disks to not be susceptible to fragmentation by GI.  This is to be
expected, since the timescales for fragmentation are so much shorter than
observed disk lifetimes/ages, it is unlikely to observe gravitationally highly unstable
PPDs.  However, we note that observed disk parameters lie close to the region of
fragmentation.

\subsection{Sensitivity to ICs}
\label{sec:SensitivityToICs}

\begin{table}
	\begin{center}
	\caption{Results for a series of tests to perturb disks near
          equilibrium.}
	\label{table:Perturbations}
	\begin{tabular}{ccccc}
	\input{\tablefolder/perturbation_tests.dat}
	\end{tabular}
	\end{center}
\end{table}

It is important to understand the causes of disk fragmentation in our
simulations.  We have paid particular care to characterize numerical issues
which may drive fragmentation.  In appendix~\ref{appendix:Convergence} we
present a simple convergence test which demonstrates that for our locally
isothermal SPH treatment, lower resolution disks fragment more easily.

To assess the sensitivity of disk fragmentation to the state of the ICs, we applied a series of small perturbations to disks close to the fragmentation boundary, with a $\Qeff$ slightly greater than 1.  We used two of the ICs in our suite of runs which didn't fragment: one with $\Qeff=1.01$ and one with $\Qeff=1.12$ (simulations 2 and 5 in table \ref{table:AllRuns}).  We found that perturbing orbital velocities slightly out of equilibrium can cause a disk to fragment, but that results were fairly insensitive to perturbing the disk height or to applying small spiral density perturbations.  These results are summarized in table~\ref{table:Perturbations}.

We applied $m = 2,3,4$ spiral density perturbations by multiplying the particle masses $M_0$ by:
\begin{equation}
M_1 = M_0 \left(1 + \epsilon \cos (m\theta)\right)
\end{equation}
where $M_1$ is the perturbed particle mass and $\epsilon$ is the depth of the perturbation, here chosen to be $\epsilon=0.01$.  These perturbations are similar to those done by e.g.~\cite{boss2002} to seed instabilities in a grid code which does not have SPH particle noise.  These perturbations were performed on the $\Qeff=1.01$ simulation and were not sufficient to force the disk to fragment.

We also wished to probe how sensitive disks are to perturbing the height.  When generating ICs, there are many different methods for estimating the vertical density profile.  We therefore wished to see whether perturbing a disk's scale height away from equilibrium could cause it to fragment.  We ran the $\Qeff=1.01$ simulation twice, multiplying the particle $z$ positions by 0.98 and 0.9, decreasing the scale heights by 2\% and 10\%, respectively.  Neither of these runs fragmented.

The disks do appear to be much more sensitive to particle velocities.  We
applied small, axisymmetric velocity perturbations to disks which otherwise did
not fragment.  The perturbed velocity $v_1$ was calculated as:
\begin{equation}
\label{eq:VelocityPerturbation}
v_1 = v_0 \left(1 + \epsilon \frac{R_0}{R}\right)
\end{equation}
where $v_0$ is the original velocity, $R_0$ is the inner edge of the disk
(where the $\Sigma$ reaches a maximum) and $\epsilon<<1$ is the depth of the perturbation.  This applies a fractional perturbation of $\epsilon$ at
$R_0$ which decays as $1/R$.  Nearly all the disk mass lies outside of $R_0$.

\begin{figure}
	 \includegraphics[width=\columnwidth]{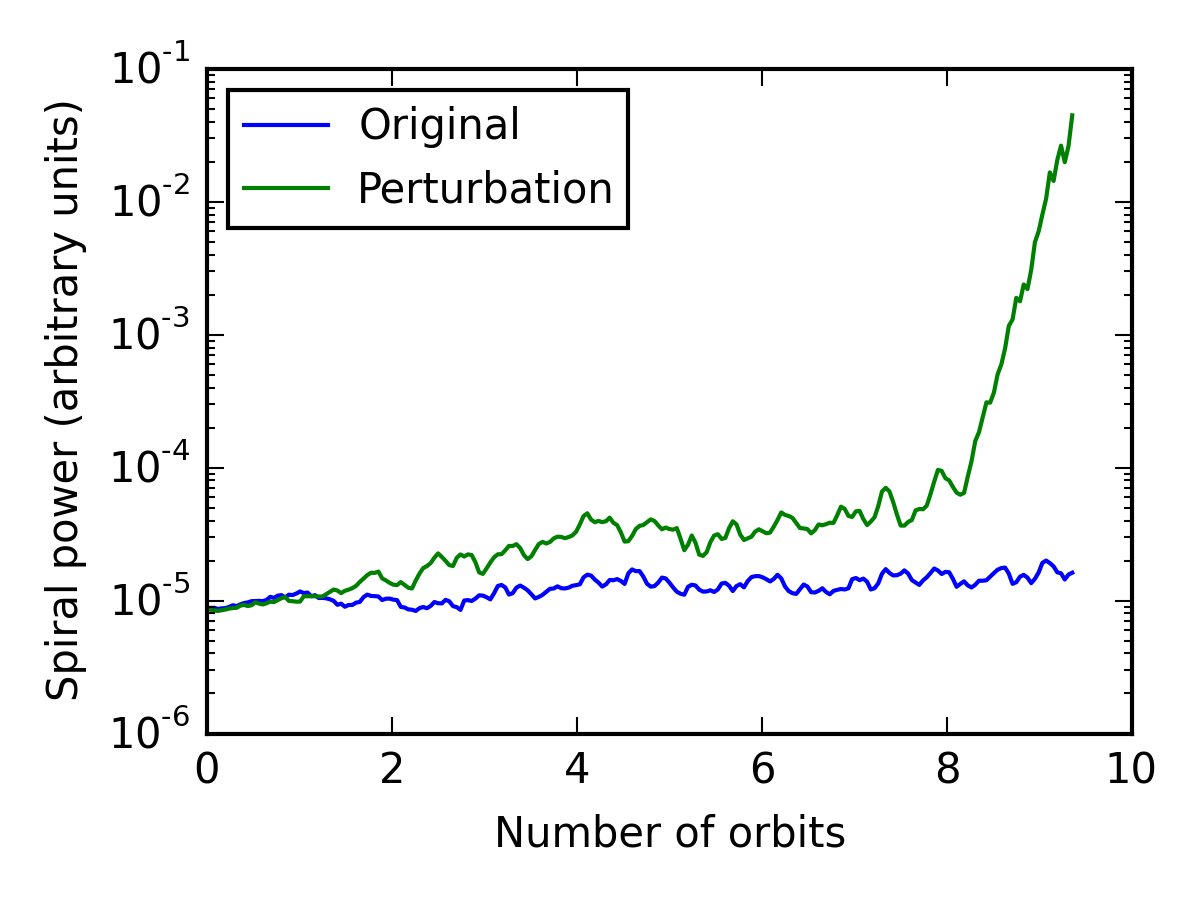}
	\caption{Total non-axisymmetric power versus time for a disk with and
		without a 1\% velocity perturbation (simulation number 2).  The original
		simulation ($\Qeff=1.01$) did not fragment, the simulation with a
		small, axisymmetric velocity perturbation does.  Spiral structure
		initially develops similarly for both simulations for about the first
		orbit.  For the next 8 orbits the perturbed simulation develops deeper
		spiral structure until it fragments after 9-10 orbits.
		\label{fig:SpiralPower}}
\end{figure}

For the $\Qeff=1.01$ disk, a 1\% perturbation ($\epsilon=0.01$) 
was sufficient to cause fragmentation.  Figure~\ref{fig:SpiralPower} shows the
spiral power as a function of time for this run.  Spiral power is calculated by binning
$\Sigma$ in $(R,\theta)$ and calculating the standard deviation.  The perturbed and
original simulations initially develop in a similar manner for the first orbital
period (defined at the most unstable radius).  During this stage an axisymmetric
($m=0$) density wave moves outward from the disk center. After about an orbital
period the perturbed disk develops significantly more pronounced spiral density
waves. After 9-10 orbits the disk fragments.  Fragmentation is accompanied by a
rapid spike in the spiral power as the disk becomes highly non-axisymmetric.

A similar test was performed with a much more stable disk ($\Qeff=1.12$).  A series of perturbations were applied ($\epsilon=.01, .02, .04, .08, .11, .16$).  An 11\% perturbation was sufficient to force the disk to fragment.  Depending on the approximations used to estimate circular velocities, discrepancies on this order can happen for disks sufficiently massive to be close to the fragmentation boundary.

This demonstrates the care which must be taken in developing equilibrium models
of unstable disks near the fragmentation boundary, especially with regard to the
velocity calculation.  An apparently small perturbation can deposit large
amounts of energy in a disk, forcing it sufficiently far of equilibrium to
fragment.

\update{
We also tested the importance of the inner and outer boundaries by taking a disk
near the fragmentation boundary with $\Qeff$ slightly above 1 and applying a step
function cut-off to $\Sigma$ on the inner edge, outer edge, and both.  Waves 
were seen to reflect off these hard boundaries, but the effect was not
sufficiently strong to drive the disk to fragmentation.
}

\section{Clumps}
\label{sec:Clumps}

\begin{figure}
	 \includegraphics[width=\columnwidth]{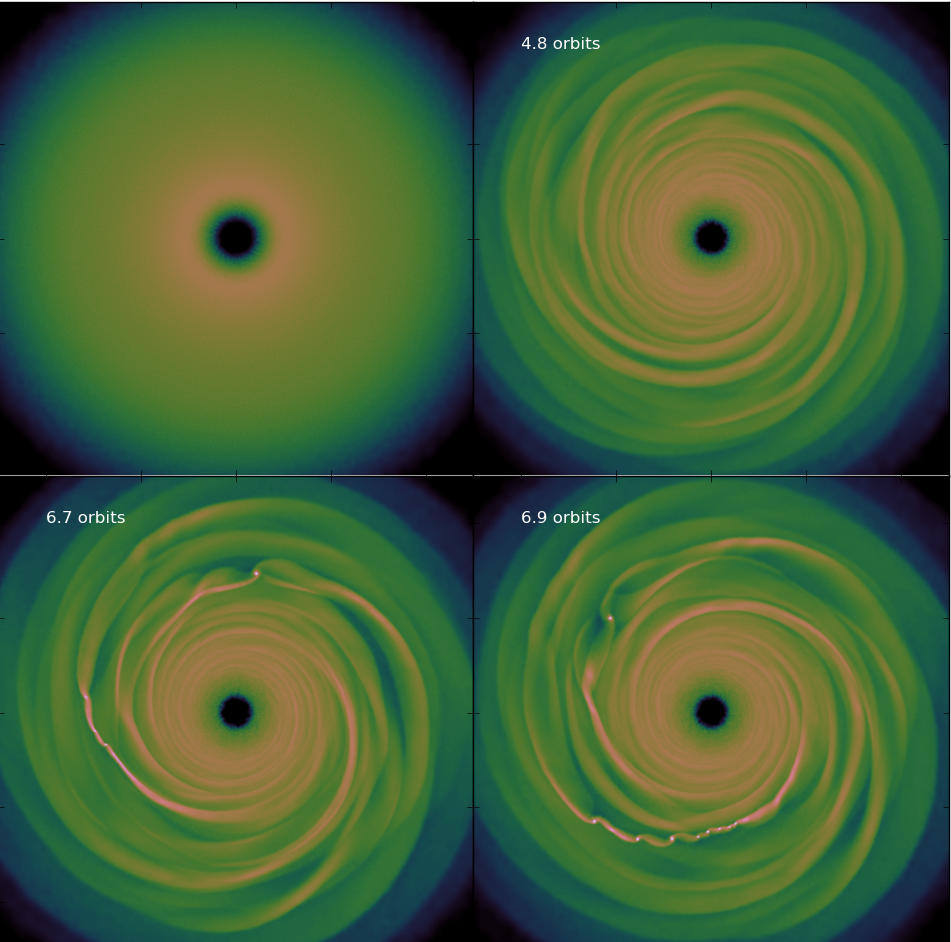}
	\caption{An example of the typical stages of clump formation.  Clockwise, from
		top-left: (1) Initial conditions $\Qeff<1$  (2) Strong spiral structure
		develops.  (3) Spiral arms become overdense and break apart into clumps. (4) The
		disk begins fragment strongly.
		\label{fig:ClumpFormation}}
\end{figure}

The formation of tightly bound, dense clumps of gas marks the stage of disk
evolution where our isothermal treatment begins to break down.  We therefore
restrict our analysis of clumps to the early stages of formation and limit the
scope of our results.

\begin{figure}
	\includegraphics[width=\columnwidth]{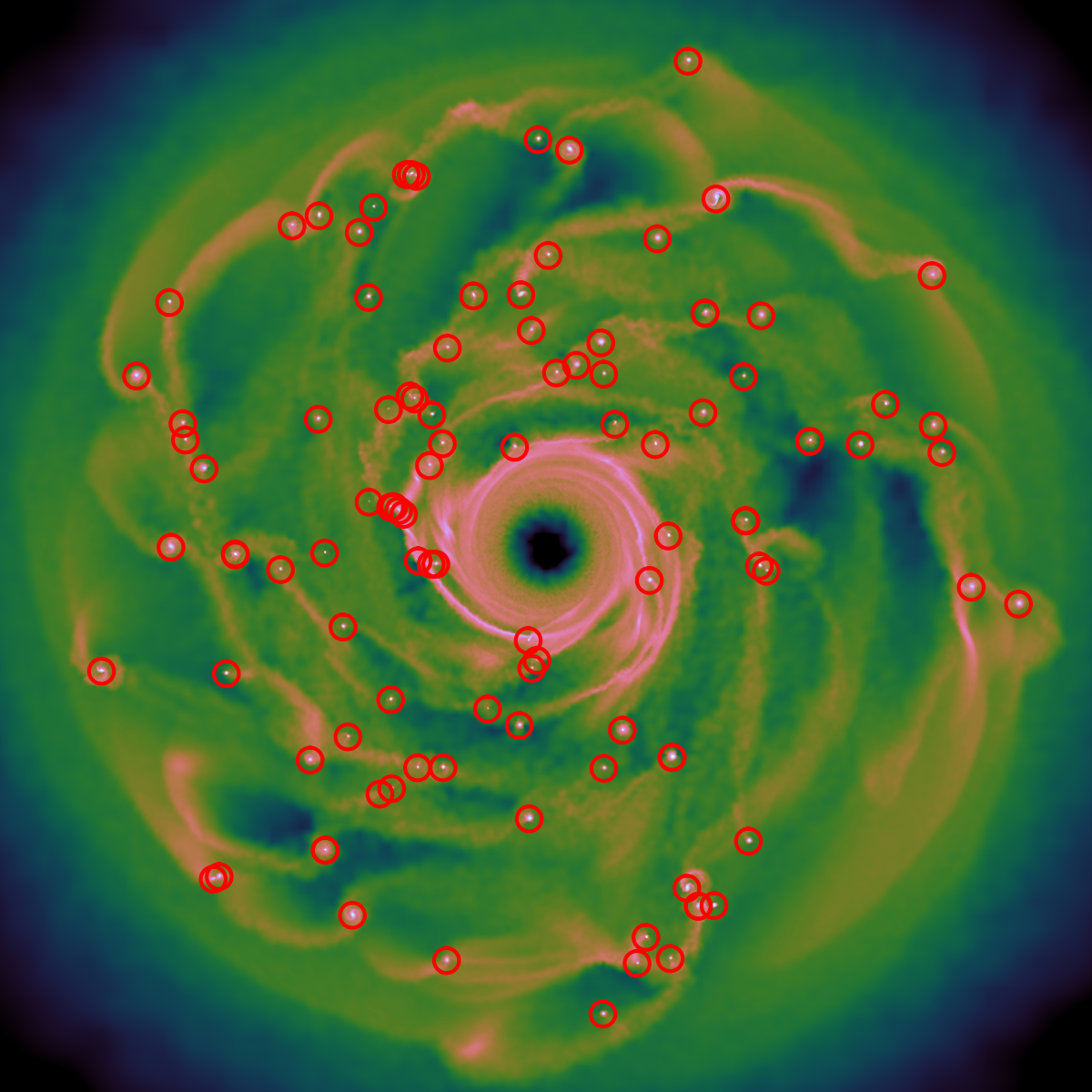}
	\caption{A demonstration of the clump finding algorithm used
		here.  Integrated column density for the gas is pictured with a logarithmic color scale.  Red
		circles mark the detected clumps.  At the end of the simulation,
		this highly unstable disk formed 108 distinct, gravitationally
		bound clumps.  The algorithm picks out clumps with a high
		success rate without reporting false positives from other high
		density structure such as spiral arms.%
		\label{fig:ClumpFind}}
\end{figure}

To track the formation of clumps, we developed a simple clump finding/tracking
software routine\footnote{Our clump finding code is freely available on github at \url{https://github.com/ibackus/diskpy} as a part of our PPD python package \textit{diskpy}} 
built around the group finder SKID\footnote{SKID is freely available at \url{https://github.com/N-BodyShop/skid}}.
To find gravitationally bound clumps, we may factor in the disk geometry. 
Particle masses are scaled by $R^3$ and a density threshold is set such that at
least $N$ particles lie within the Hill-sphere of particles under consideration,
where $N$ is chosen as the number of neighbors used for SPH smoothing.
This gives a threshold of 
\begin{equation}
\label{eq:DensityThreshold}
\rho _{min} = \frac{3N\MStar}{R^3}
\end{equation}  The SKID algorithm is then applied, which uses a
friends-of-friends clustering algorithm followed by a gravitational unbinding
procedure to determine which clumps are gravitationally bound.  By visual
inspection, this provides robust results over a large range of disk and
simulation parameters, and importantly avoids marking high-density spiral arms
as clumps.  Figure~\ref{fig:ClumpFind} shows an example of this clump finding
applied to a highly unstable PPD.

To track clumps over many time steps, clumps are first found in all simulation
snapshots.  They are then tracked over time by comparing clumps in adjacent time
steps and seeing which have the most particles in common.  Mergers (including
multiple mergers), clump destruction, and clump formation are accounted for. 
Clump parameters such as mass, density, size, and location are all calculated
and followed as a function of time.

We find that clumps form according to the same general picture, as shown in
figure~\ref{fig:ClumpFormation}.  Disks with a minimum $\Qeff \lesssim 1.6$ (or
about $Q_{min} \lesssim 1.3$) will form noticeable spiral structure after
several orbital periods.  Disks with an initial $Q_{eff} < 1$ will grow
overdense spiral arms which will collapse and fragment into several dense
clumps.  Clumps initially form near the most unstable disk radius.  For these
simulations, that is approximately at $R_d$.  They form after several to tens of
orbital periods at the most unstable radius.  Average clump masses are around
$0.3 M_{Jupiter}$, with some rapidly growing to $1 M_{Jupiter}$. The timescales for disk
fragmentation increase rapidly as $\Qeff\rightarrow 1$ (fig.~\ref{fig:Runtime}).
After this stage, the isothermal approximation begins to break down.  The disks
then undergo a rapid, violent fragmentation.  

\section{Discussion}%
\label{sec:Discussion}

\subsection{Thermodynamics}
\label{sec:ThermodynamicsDiscussion}

For these simulations we used a locally isothermal approximation for several
reasons.  We wished to perform a large scan of parameter space without
compromising resolution too strongly.  A computationally fast isothermal EOS is
straightforward to implement.  We also desired to build on previous work and to 
extend it to poorly studied M-dwarf systems.  Our work here is directed at	
exploring the dependence of the fragmentation boundary on stellar/disk mass,
disk height, and ICs.  We leave the dependence on EOS for future work.

\update{
A non-isothermal EOS introduces non-trivial numerical issues, especially in the
context of SPH simulations of protoplanetary disks.  Unwarranted, poorly
understood heating terms, especially from artificial viscosity (AV), are
introduced into the energy equation.  Previous results
\citep{meru2011,meru2012,lodato2011,rice2012,rice2014} and our own initial tests indicate that in
the context of PPDs, AV heating can dominate disk thermodynamics.  Non-isothermal PPD simulations may not converge \citep{meru2011}, whereas in Appendix~\ref{appendix:Convergence} we demonstrate that our approach does converge.  It is also
unclear how rapidly disks will radiatively cool, an important parameter for the
possibility of disk fragmentation \citep{gammie2001,rafikov2005,rafikov2007}. 
We hope to investigate these effects in future work.
}

The isothermal approximation used for these simulations limits the scope of our
results.  An isothermal EOS would well
approximate a disk where stellar radiation, viscous accretion heating, and
radiative losses to infinity, are nearly balanced and control the temperature of
the disk.  Furthermore, temperature is independent of density,
which is only appropriate for an optically thin disk.  The isothermal
approximation applies only to scenarios where dynamical
timescales are much longer than the timescales for heating/cooling back to
thermal equilibrium with the background.

\update{
These conditions may not hold in the disks under consideration.  This experiment
therefore does not realistically follow the thermal evolution of the disk.  We
are limited to a preliminary investigation into the large-scale dynamics of the
disk before the putative equilibrium temperature profile would be expected to be
strongly altered.

During the initial stages of disk evolution, dynamical and physical timescales
are of order the orbital period and disk radius, respectively.  During this
stage, the isothermal EOS can still provide insight into the  global dynamics of
a GI disk at a certain stage. However, once clumps form, the isothermal
approximation no longer provides much insight.  Clumps should get hot as they
collapse.  The dynamic timescales of dense clumps will be short as they accrete
matter, decouple from the disk, and scatter with other clumps.  Pressure support
of clumps, which is poorly captured by an isothermal EOS,
}
should tend to increase
their size and their coupling to the disk, meaning that the
violent fragmentation of disks after initial clump formation which we see in our
simulations may not be the final state of a typical fragmenting disk.  Clumps
which are sufficiently dense may decouple from the disk enough to experience
strong shocks and tidal interactions which will cause heating.  These processes
will strongly influence clump growth, evolution, and survival, all of which are
still under investigation \citep{nayakshin2010,galvagni2012}.

We therefore limit ourselves to discussing the early stages of clump formation.
Our results indicate that the critical value of $Q_{min} \lesssim 0.9$ required for fragmentation is
significantly lower than some previous results which found closer to
$Q_{min}\lesssim 1.5$.  Although this makes requirements for fragmentation more
stringent, it does not rule out GI and disk fragmentation as important
mechanisms during planet formation in PPDs around M-dwarfs.

\subsection{Previous results}
\label{sec:ICDiscussion}

We find that for most disks, $Q \lesssim 0.9$ is required for disk
fragmentation.  Re-parameterizing $Q$ as $Q_{eff}$ to include the stabilizing
effect of disk height provides a more precise way to predict fragmentation.  The
ratio of $Q_{eff}/Q$ can vary by 30\% for reasonable disk parameters.  This
ratio will vary even more when considering solar type stars in addition to
M-dwarfs.  However, $Q$ still provides a reasonable metric for disk
fragmentation.

\update{Other isothermal studies have found different boundaries.  In contrast to our results, \cite{nelson1998} found the threshold to be $Q\leq1.5$.}  \cite{boss1998} found $Q$ as high as 1.3 would
fragment.   \cite{boss2002} found, using an isothermal EOS or diffusive
radiative transfer, that $Q=1.3-1.5$ would fragment.  \cite{mqws} found $Q=1.4$
isothermal disks would fragment.  \cite{pickett2003} found that cooling a disk
from $Q=1.8$ to $Q=0.9$ caused it to fragment.  Their clumps did not survive, although as
they note that may be due to numerical issues.  \cite{boley2009} found that
disks could be pushed below $Q=1$ by mass loading and still not fragment, by
transporting matter away from the star and thereby decreasing $\Sigma$ and
increasing $Q$.  \update{It should be noted that some of these simulations were run at much lower resolution than ours.  Differences may also be due in part to simulation methods: using cylindrical grids, spherical grids, or SPH methods; applying perturbations; or even 2D \citep{nelson1998} vs 3D simulations.}

Since all the details of previous work are not available, the source of the
discrepancy in critical $Q$ values is uncertain.  One source of discrepancy is
simply how $Q$ is calculated.  As mentioned in
\S\ref{sec:FragmentationAnalysis}, approximating $Q$ by ignoring disk
self-gravity and pressure forces can overestimate $Q$ on the 10\% level for
heavy disks.

\begin{figure}

	\includegraphics[width=\columnwidth]{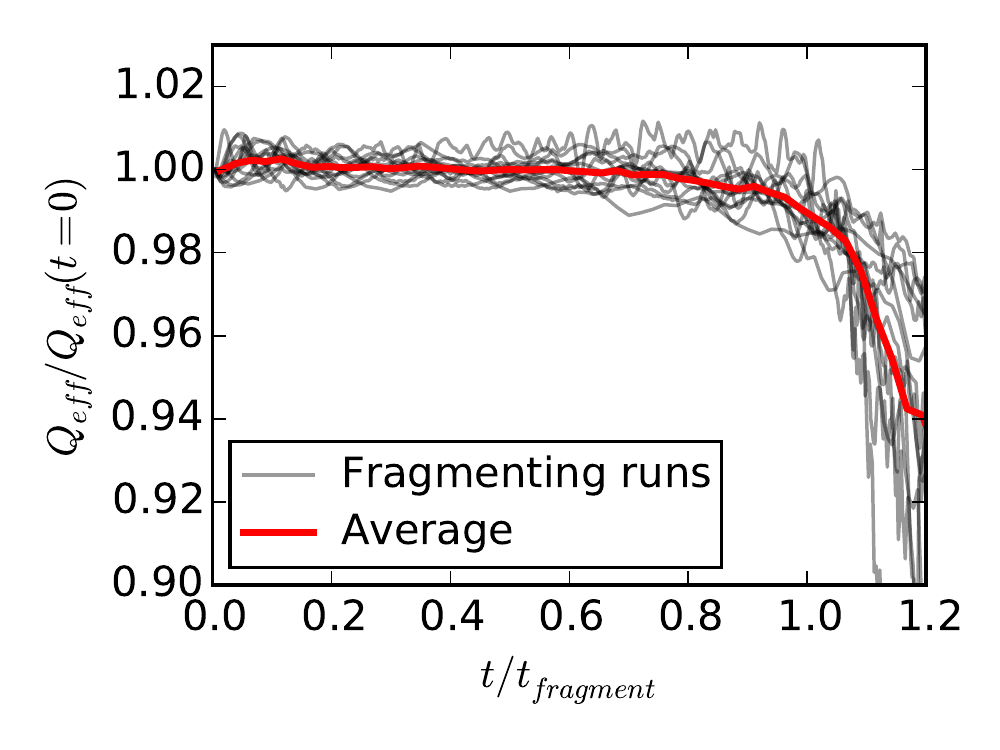}
	\caption{\secondupdate{
	Minimum \mbox{$\Qeff$} normalized by the initial minimum \mbox{$\Qeff$} vs fraction of time until fragmentation for all the fragmenting runs.  The average of these runs is plotted in red.  Disk fragmentation occurs at \mbox{$t/t_{fragment}=1$}.  All runs follow similar trajectories in this plot, even though a significant range of initial \mbox{$\Qeff$} and \mbox{$t_{fragment}$} values are represented here (all the fragmenting runs in Fig.~\ref{fig:Runtime} are presented here).  The simulations undergo an initially gradual decrease in \mbox{$\Qeff$} which steepens sharply shortly before fragmentation.} %
		\label{fig:QeffvsTime}}

\end{figure}

The discrepancy may also be due to the different methods of constructing
equilibrium disks.  As demonstrated in \S\ref{sec:SensitivityToICs},
overestimating velocities at less than the percent level can force a disk to
fragment.  Disks near the fragmentation boundary are very sensitive to ICs. 
\secondupdate{Initial conditions are not in general available for previous work,
however we can note that some studies appear to display a rapid evolution of $Q$
at the beginning of the simulation.  For example, some runs of
\cite{pickett2003} evolve from $Q_{min}=1.5$ to $Q_{min}=1$ in fewer than 3~ORPs
(see their Figure~14).  Figure~1 of \cite{mqws} shows two isothermal simulations
which evolve from a $Q_{min}$ of 1.38 and 1.65 to $Q_{min}=1$ in fewer than
2~ORPs.  This is indicative of ICs which are out of equilibrium.

In contrast, even our very unstable disks display a remarkably smooth and
gradual initial evolution.  Figure~\ref{fig:QeffvsTime} shows the behavior of
the minimum $\Qeff$ (normalized by its initial value) for our fragmenting runs
as a function of time until fragmentation.  For all the runs, $\Qeff$ decreases
gradually for most of the simulation until dropping rapidly shortly before the
disk fragments. For our runs near the fragmentation boundary, this is much more gradual and much less pronounced than for the runs of
\cite{pickett2003} or \cite{mqws} mentioned above.  For us, a much smaller change in $Q$ takes around 10~ORPs.   $\Qeff$ evolves even more
slowly for non-fragmenting runs.  $Q_{min}$ follows a similar behavior, although
with more scatter (in large part because $Q$ does not determine the timescale
until fragmentation as well as $\Qeff$ does).}

However it is not certain how close to equilibrium ICs should be to capture the
relevant physics of PPDs.  Actual disks are constantly evolving from the early
stages of star formation until the end of the disk lifetime.  We chose to use
disks as close to equilibrium as possible, seeded only with SPH poisson noise,
to avoid introducing numerical artifacts.  Some authors introduce density
perturbations which are controllable.  If sufficiently large, they may serve to
ameliorate the problems mentioned above by explicitly having fragmentation be
driven by physically reasonable spiral modes (e.g.~\cite{boss2002}) or
intentionally large random perturbations (e.g.~\cite{boley2009}).  Others have
considered mass loading as a means to grow to low $Q$ \citep{boley2009}.

\subsection{GI in PPDs}

Observed disks around M-dwarfs do not appear to have low enough inferred $Q_{min}$ values
to be sufficiently unstable for fragmentation under GI, however this is what
would be expected given the short timescales for the fragmentation of an
unstable disk.  Reported disk ages are of order $10^6$~years \citep{haisch2011}, orders of magnitude
longer than fragmentation timescales, making the observation of a highly 
gravitationally unstable disk unlikely.  This is a strong selection effect for
observed disk parameters.

Although fragmentation timescales are very rapid, disks may persist much longer
in moderately unstable configurations where GI drives large scale structure but
which have not grown sufficiently unstable as to be prone to fragmentation.  Early work is being
done on trying to observe GI driven structures, but with current instrumentation
such structures will be difficult to resolve \citep{douglas2013,evans2015}.

As shown in figure \ref{fig:FragBound}, the fiducial values for disk parameters
adopted here place observed disks near the boundary for fragmentation.  The fact
that observations indicate normal disk parameters which are close to the
boundary, rather than orders of magnitude off, suggests it is plausible that a
significant portion of PPDs around M-dwarfs will undergo fragmentation.  This
would predict a sharp transition in the distribution of inferred $\Qeff$ values around $\Qeff=1$.  Older disks tend to expand radially
\citep{isella2009}, thereby decreasing $\Sigma$, increasing $Q$, and pushing
them away from the $\Qeff=1$ boundary.  We note that while $Q$ is a reasonably
strong predictor of fragmentation, disk height is an additional parameter worth
measuring to predict fragmentation.

Given the results of these isothermal simulations, we expect GI to play a large
role in the early stages of planet formation \update{ around M-dwarfs.  
The exact role of star mass/type on fragmentation remains unclear.
The parameter space we scanned is sufficiently large that adding an
extra dimension was prohibitive, we therefore only studied one star mass.  Future work should be able to determine what stars are the most suitable for fragmentation.
}  Once large-scale density
perturbations are formed via GI, the fate of the disk remains unclear.  Future
work should include more sophisticated thermodynamics to follow the evolution of
the gaseous component of the disk to better determine under what conditions
clumps will form, and what is required for them to survive.  Additionally, decreasing the resolution in our isothermal SPH simulations appears to drive fragmentation (see~\S\ref{appendix:Convergence}).  The importance of resolution in simulations is subject to much investigation and should be further pursued (e.g.~\cite{meru2011,meru2012,cartwright2009}).

Planet formation will of course require the concentration of solids as well.
Including dust in simulations of young PPDs will be required.  In a fully 3D,
highly non-axisymmetric environment, we may investigate how GI affects solids.
Dust enhancement through pressure gradients, dust evolution through collisions
and coagulation, and dust coupling to disk opacity and cooling, will all
strongly affect prospects for planet formation.  Dust dynamics in PPDs using SPH
has been proposed recently \citep{price2012a,price2015}, and we hope to explore
solid/gas interactions in the future.

\section*{Code and data}
The code repository for generating our ICs and for clump-tracking is freely available online at \mbox{\url{https://github.com/ibackus/diskpy}} as a part of our \textit{diskpy} python package.  IC generation is contained within the subpackage \textit{diskpy.ICgen} and clump tracking is contained within the subpackage \textit{diskpy.clumps}.  A public version of ChaNGa is available at \url{http://www-hpcc.astro.washington.edu/tools/changa.html} and is required for IC generation.  The group finding software SKID (required for clump finding) is available at \url{https://github.com/N-BodyShop/skid} and depends also on tipsy tools (\url{https://github.com/N-BodyShop/tipsy_tools}).

We have also made much of our data available online at the University of
Washington's ResearchWorks archive at \url{http://hdl.handle.net/1773/34933}. 
Initial conditions and final simulation snapshots are available for all the
simulations presented here.  Wengen test results (see
appendix~\ref{appendix:Wengen}) are also available there.

\section*{Acknowledgements} 
We would like to acknowledge Aaron Boley for many fruitful discussions and much feedback on this work.  We thank David Fleming for help on this work and \textit{diskpy}.
This work was performed as part of the NASA Astrobiology Institute's
Virtual Planetary Laboratory, supported by the National Aeronautics
and Space Administration through the NASA Astrobiology Institute under
solicitation NNH12ZDA002C and Cooperative Agreement Number
NNA13AA93A.  The authors were also supported by NASA grant NNX15AE18G.
This work used the Extreme Science and Engineering Discovery
Environment (XSEDE)\citep{XSEDE}, which is supported by National
Science Foundation grant number ACI-1053575.
This work was facilitated though the use of advanced computational, storage, and networking infrastructure provided by the Hyak supercomputer system at the University of Washington.

\bibliographystyle{mnras}
\bibliography{mybib}

\appendix

\section{Wengen tests}
\label{appendix:Wengen}

The Wengen tests are a series of code tests designed to compare different
astrophysical hydrodynamic and gravity simulation codes and are available online
at \mbox{\url{http://www.astrosim.net/code/}}.  We ran Wengen test 4, an
unstable isothermal PPD, with ChaNGa.  We present the results of our simulation
here.  We find that our results are in good agreement with other simulation
codes (previous test results can be found at
\mbox{\url{http://users.camk.edu.pl/gawrysz/test4/}}).  We have reproduced all
the plots on the Wengen test website for the ChaNGa results, which are available
at \url{http://hdl.handle.net/1773/34933}.  Our ICs are the 200k-particle run.  Here we
present a few figures demonstrating the ChaNGa results.

\begin{figure}
	 \includegraphics[width=\columnwidth]{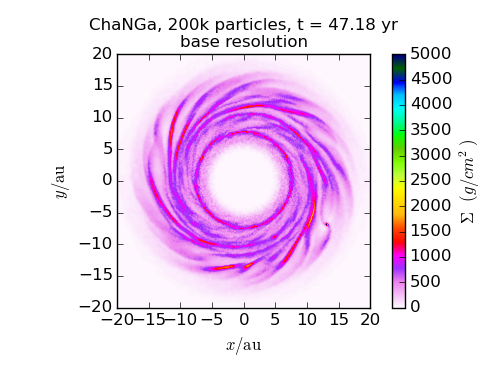}
	\caption{Surface density map of the Wengen test simulation at $t=47.18$ years (6 in code units).  This reproduces the \textit{surface density map} at $t=6$ in the images table at \protect\url{http://users.camk.edu.pl/gawrysz/test4/\#images}.   As with the other SPH codes and the higher resolution runs, a clump has formed.
		\label{fig:Wengen:SurfaceDensity}}
\end{figure}

\begin{figure}
	 \includegraphics[width=\columnwidth]{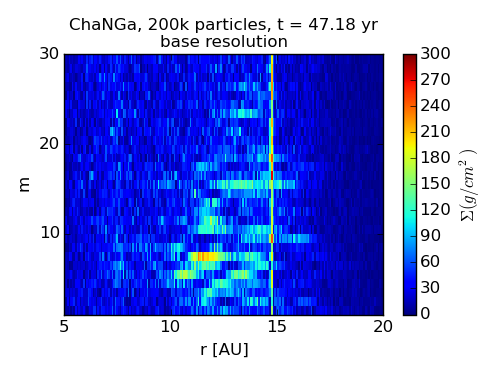}
	\caption{Amplitude of the fourier transform of the surface density along the angular direction as a function of radius, for the same snapshot in figure~\ref{fig:Wengen:SurfaceDensity}.  This reproduces the \textit{FFT of surface density} plots at $t=6$ in the images table at \protect\url{http://users.camk.edu.pl/gawrysz/test4/\#images}.  The clump which has formed shows up as a bright vertical stripe.  200 radial bins were used.  The features in the plot agree well with those for other SPH codes and the high resolution codes.%
		\label{fig:Wengen:FFTSurfaceDensity}}
\end{figure}

\begin{figure}
	 \includegraphics[width=\columnwidth]{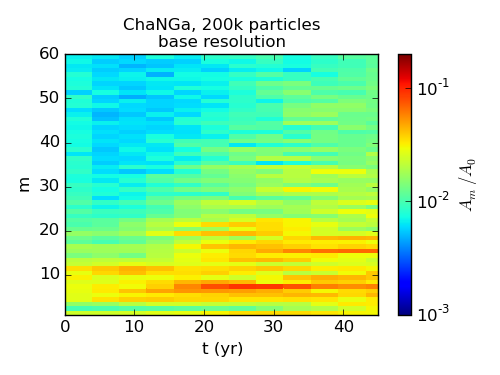}
	\caption{Radially integrated fourier transform of surface density as a function of time.  The amplitudes are normalized by the DC amplitude (not shown).  This reproduces the \textit{FFT integrated over r}, $A_m (t)$ plots in the images tables available at \protect\url{http://users.camk.edu.pl/gawrysz/test4/\#images}.  The development of strong power around $m=8$ and later around $m=16$ agrees with other codes.
		\label{fig:Wengen:PowerVsTime}}
\end{figure}

\section{Convergence test}
\label{appendix:Convergence}

Previous work has indicated that the results of SPH simulations of PPDs can be
resolution dependent \citep{lodato2011,meru2011,meru2012}.  
\secondupdate{\cite{nelson2006} laid out several resolution requirements for SPH
simulations of PPDs.  For our suite of simulations, we exceed the mass
resolution requirement by a minimum factor of 7, and most simulations exceed it
by a factor of 40.  We also easily meet their scale height resolution
requirement.  At the midplane of the most unstable disk radius, the ratio of the
smoothing length to the scale height is between $4-12$ for all our runs
(\cite{nelson2006} finds this ratio should be at least $\sim4$).
}

As part of our analysis, we ran a basic convergence test to verify our simulation
code and to investigate the effects of resolution on disk fragmentation in SPH
simulations.   We ran a simulation close to the fragmentation boundary with a
minimum $\Qeff=1.06$ (simulation 48 in table~\ref{table:AllRuns}) at 6 different
resolutions from 50k to 10M particles.  These runs are summarized in
table~\ref{table:Convergence}.

We find (i) that low resolution simulations are more susceptible to disk
fragmentation and (ii) that simulations appear to converge reasonably well, with
our chosen resolution of $10^6$-particles being sufficient for the analysis
presented in this paper.  However, these convergence tests are only preliminary
and future work may reveal that higher particle count is required for fully
believable results.  \update{Convergence may also depend on EOS, cooling prescriptions, and whether a grid code or an SPH code is used.}

\begin{figure*}
	\includegraphics[width=\textwidth]{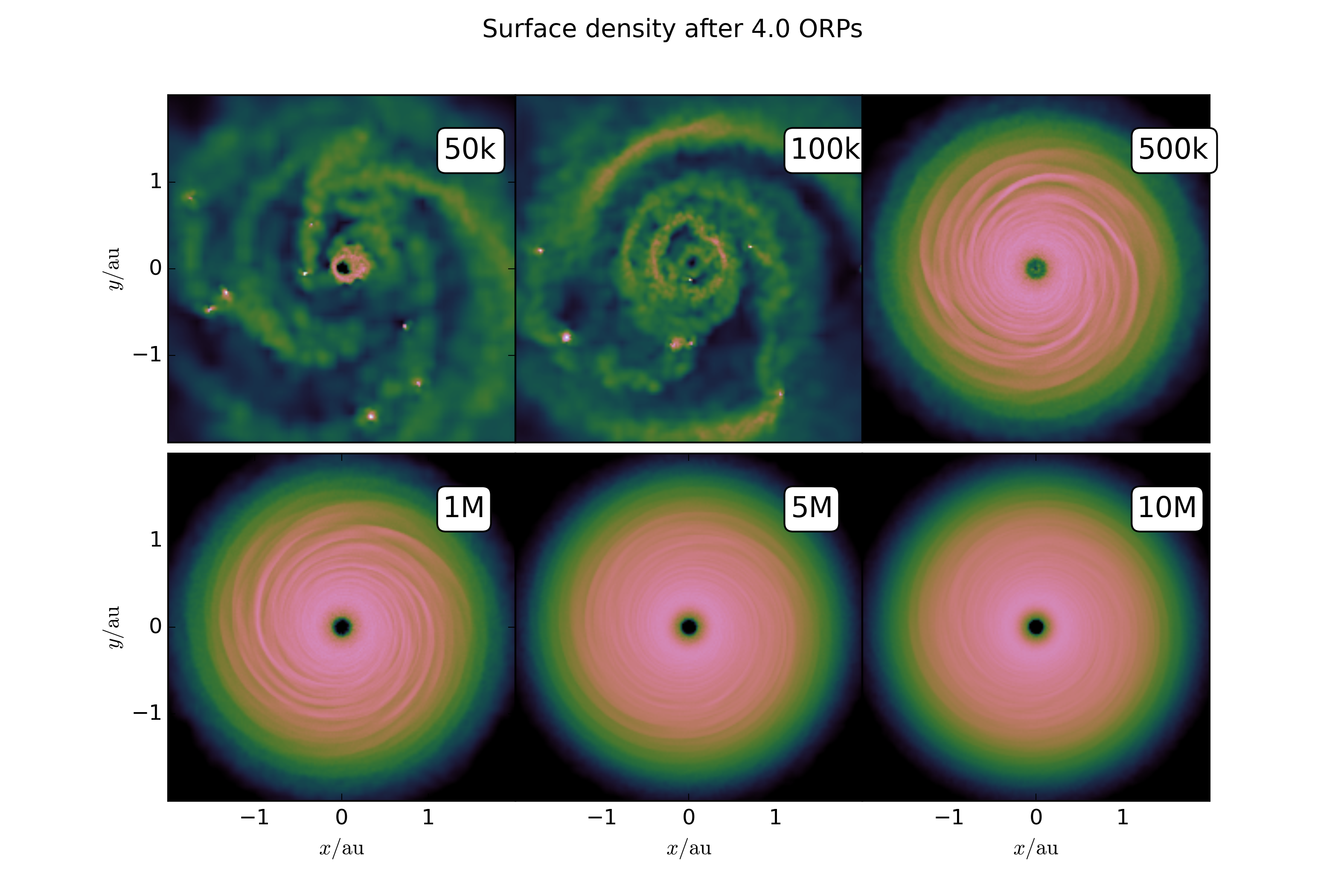}
	\caption{Logarithmic surface density plots for the simulations listed in table~\ref{table:Convergence} of a $\Qeff=1.06$ run after 4.0 ORPs (defined at the initial disk radius).  The low resolution runs have already fragmented.  The 500k-particle run has developed strong spiral power and will eventually fragment.  The remaining runs do not fragment and approach a value of $\Qeff > 1.1$ within the course of the simulations.
		\label{fig:Convergence:AllRuns}}
\end{figure*}

Figure~\ref{fig:Convergence:AllRuns} shows logarithmic surface density plots of
all 6 runs after 4.0~ORPs.  The 50k- and 100k-particle runs have already
fragmented violently.  The 500k-particle run has developed somewhat stronger
spiral power than the higher resolution runs and eventually fragments after
about 12~ORPs. The other runs have developed some spiral power that is
insufficient to drive the disks to fragmentation.

\begin{figure}
	\includegraphics[width=\columnwidth]{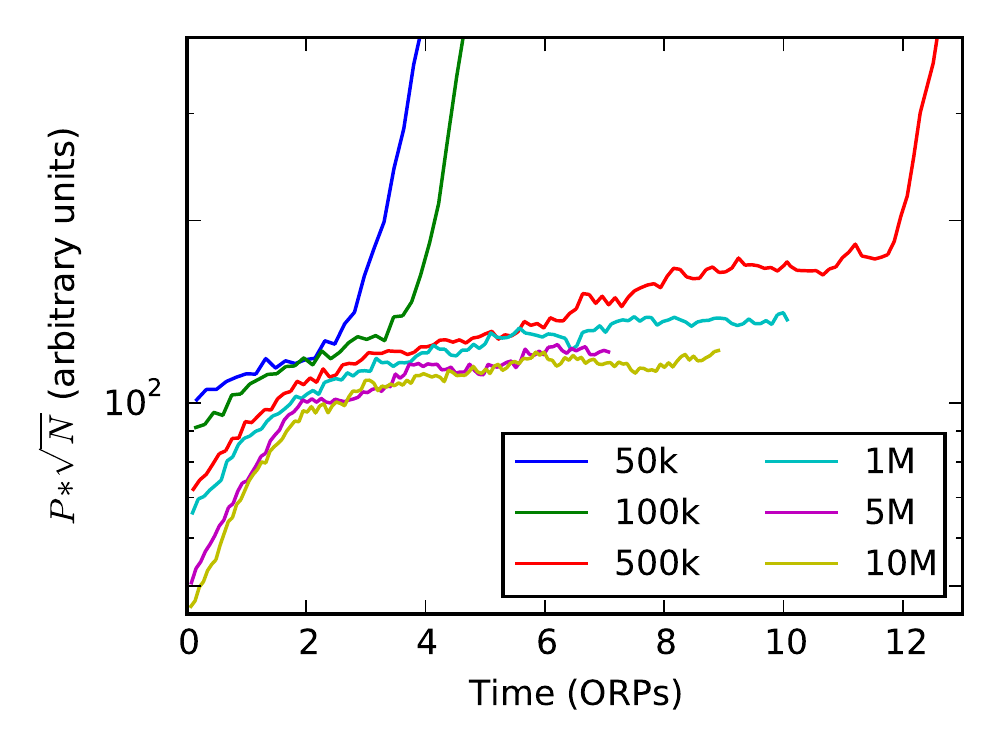}
	\caption{Normalized spiral power vs. time for the 6 simulations in
		the convergence test.  Power is calculated by binning $\Sigma$ in
		$(R,\theta)$, calculating the standard deviation along $\theta$ and
		summing along $R$.  Power is then normalized by multiplying by $\sqrt{N}$ to
		adjust for the power in particle noise.  Fragmentation is seen for
		the simulations with 50k-, 100k-, and 500k-particles as a rapid increase
		in spiral power at the end of the simulation.  As expected, the higher resolution simulations do not fragment.%
		\label{fig:Convergence:PowerVsTime}}
\end{figure}

Figure~\ref{fig:Convergence:PowerVsTime} shows what we call the normalized
spiral power.  We calculate spiral power by binning the surface density in
$R,\theta$ (we used $128\times128$ bins here), calculating the standard
deviation along the angular direction, and summing along the radial direction. 
We then normalize by multiplying the spiral power by $\sqrt{N}$, where $N$ is
the number of SPH particles in the run.  This is done to account for noise in
the number of particles per bin which scales as $\propto 1/\sqrt{n_{per~bin}}$
where $n_{per~bin}$ is approximately proportional to $N$.

The normalized spiral
power then represents how much larger than the particle noise the spiral power
is.  Fragmentation is visible in figure~\ref{fig:Convergence:PowerVsTime} as a
sharp rise in power at the end of the simulation.  As can be seen, the
normalized spiral power decreases with increasing particle count and converges
for the higher-resolution simulations.

As expected for $\Qeff=1.06$, the higher resolution simulations do not fragment.
As shown in table~\ref{table:Convergence}, the simulations with $N\geq10^6$
particles approach a stable value of $\Qeff\approxeq1.11$ and therefore would
not fragment if run for longer.  Simulations with $N\leq500\mathrm{k}$ particles
approached very low $\Qeff$ minimum values, although $\Qeff$ is strictly
speaking not well defined for highly non-axisymmetric disks.

\begin{table}
\begin{center}
\caption{Table of convergence tests runs.  ICs are identical to simulation 48 in table~\ref{table:AllRuns} but with a different number of particles.  The resolution is the number of SPH particles in the run.  Runs that fragment are highly non-axisymmetric, so the quoted $\Qeff$ values are only illustrative.  Runs that don't fragment approach a stable value significantly above 1.}
\label{table:Convergence}
\begin{tabular}{c c c c}
Run & Resolution & Fragment? & Final $Q_{eff}$\\
\hline
0 & 50k & yes & 0.67 \\
1 & 100k & yes & 0.69 \\
2 & 500k & yes & 0.89 \\
3 & 1M & --- & 1.11 \\
4 & 5M & --- & 1.11 \\
5 & 10M & --- & 1.12 \\
\end{tabular}
\end{center}
\end{table}

\onecolumn
\begin{table}
\caption{The suite of runs presented here.}
\label{table:AllRuns}
\begin{supertabular}{ccccccccc}
	\input{\tablefolder/table.dat}
\end{supertabular}
\end{table}

\bsp	
\label{lastpage}
\end{document}